\documentclass[floats,floatfix,amssymb,prd,twocolumn,nofootinbib,superscriptaddress]{revtex4-1}
\usepackage{amssymb,amsmath,verbatim,mathtools,needspace,enumitem,etoolbox,graphicx,physics,microtype}

\usepackage[dvipsnames, usenames]{xcolor}

\definecolor{linkcolor}{rgb}{0.0,0.3,0.5}
\definecolor{urlcolor}{rgb}{0.27,0.55,0.}
\definecolor{funcolor}{rgb}{0.65, 0.16, 0.16}
\usepackage[all]{hypcap}
\usepackage{stackengine,xcolor,scalerel,xspace,multirow}

\usepackage[T1]{fontenc}
\usepackage[utf8]{inputenc}
\usepackage[normalem]{ulem}

\allowdisplaybreaks
\def\apj{\rm{ApJ}}
\def\apjl{\rm{ApJ}}
\def\apjs{\rm{ApJS}}
\def\aap{\rm{A\&A}}

\def\mnras{\rm{MNRAS}}
\def\prd{\rm{PRD}}
\def\prl{\rm{PRL}}
\def\prx{\rm{PRX}}
\def\nat{\rm{Nature}}
\def\cqg{\rm{CQG}}

\def\lrr{\rm{LRR}}

\newcommand{\chieff}{\ensuremath{\chi_{\mathrm{eff}}}\xspace}
\newcommand{\chieffMedian}{Median of \chieff \xspace}

\newcommand{\msun}{\ensuremath{M_{\odot}}\xspace}

\newcommand{\beqn}{\begin{eqnarray}}
\newcommand{\enqn}{\end{eqnarray}}
\newcommand{\linf}{\textsc{LALInference}}

\newcommand{\fa}{\ensuremath{f_a}\xspace}
\newcommand{\engineurl}{\url{superstring.mit.edu/welcome.html}\xspace}

\begin{document}

\title{Gravitational-wave astrophysics with effective-spin measurements:  Asymmetries and selection biases}

\author{Ken K.~Y. Ng}
\email{kenkyng@mit.edu}
\affiliation{LIGO, Massachusetts Institute of Technology, Cambridge, Massachusetts 02139, USA}
\affiliation{Department of Physics and Kavli Institute for Astrophysics and Space Research, Massachusetts Institute of Technology, Cambridge, Massachusetts 02139, USA}

\author{Salvatore Vitale}
\affiliation{LIGO, Massachusetts Institute of Technology, Cambridge, Massachusetts 02139, USA}
\affiliation{Department of Physics and Kavli Institute for Astrophysics and Space Research, Massachusetts Institute of Technology, Cambridge, Massachusetts 02139, USA}

\author{Aaron Zimmerman}
\author{Katerina Chatziioannou}
\affiliation{Canadian Institute for Theoretical Astrophysics, 60 St.~George Street, University of Toronto, Toronto, Ontario M5S 3H8, Canada}

\author{Davide Gerosa}
\thanks{Einstein Fellow}
\affiliation{TAPIR 350-17, California Institute of Technology, 1200 E California Boulevard, Pasadena, California 91125, USA}

\author{Carl-Johan Haster}
\affiliation{Canadian Institute for Theoretical Astrophysics, 60 St.~George Street, University of Toronto, Toronto, Ontario M5S 3H8, Canada}

\date{\today}

\begin{abstract}
Gravitational waves emitted by coalescing compact objects carry information about the spin of the individual bodies. However, with present detectors only the mass-weighted combination of the components of the spin along the orbital angular momentum can be measured accurately. 
This quantity, the effective spin \chieff, is conserved up to at least the second post-Newtonian order.
The measured distribution of \chieff values from a population of detected binaries, and in particular whether this distribution is symmetric about zero, encodes valuable information about the underlying compact-binary formation channels.
In this paper we focus on two important complications of using the effective spin to study astrophysical population properties: (i) an astrophysical distribution for \chieff values which is symmetric does not necessarily lead to a symmetric distribution for the detected effective spin values, leading to a \emph{selection bias}; and (ii) the posterior distribution of \chieff for individual events is \emph{asymmetric} and it cannot usually be treated as a Gaussian. 
We find that the posterior distributions for \chieff systematically show fatter tails toward larger positive values, unless the total mass is large or the mass ratio $m_2/m_1$ is smaller than $\sim 1/2$. 
Finally we show that uncertainties in the measurement of \chieff are systematically larger when the true value is negative than when it is positive.
All these factors can bias astrophysical inference about the population when we have more than $\sim 100$ events and should be taken into account when using gravitational-wave measurements to characterize astrophysical populations.

\end{abstract}


\maketitle

\section{Introduction}
With the first two scientific runs of the advanced LIGO and VIRGO detectors~\cite{2015CQGra..32g4001L,2015CQGra..32b4001A} now completed, observations of gravitational waves (GWs) emitted by coalescing binary black holes (BBHs) and binary neutron stars (BNSs) are becoming routine \cite{2016PhRvL.116f1102A,2016PhRvL.116x1103A,2017PhRvL.118v1101A,2017ApJ...851L..35A,2017PhRvL.119n1101A,2017PhRvL.119p1101A}.
Rates inferred for the merger of compact objects imply that dozens of sources will be detected every year by current ground-based detectors at design sensitivity~\cite{2016PhRvX...6d1015A,2017PhRvL.119p1101A,2016ApJ...832L..21A,2016ApJ...833L...1A}.
Such rates will allow us to move beyond characterizing individual objects to characterizing whole populations, revealing details about the underlying astrophysics of compact binaries.

The spins of the two merging objects are among the cleanest indicators of the underlying formation channels (though others have been proposed, e.g.~the orbital eccentricity; see Refs.~\cite{2016PhRvD..94f4020N,2017MNRAS.465.4375N,2016ApJ...830L..18B,2018arXiv180208654S,2018arXiv180406519S}).
In fact, the main proposed formation pathways for compact-binary coalescences (CBCs) result in different distributions for the orientations of the component spins~\cite{2010CQGra..27k4007M}. Systems formed via dynamical interactions in globular clusters \cite{2013LRR....16....4B} or stellar clusters near active galactic nuclei~\cite{2009ApJ...692..917M} are typically expected to have a random distribution of the spins' angles.
Conversely, binaries formed through common envelope evolution in galactic fields \cite{2014LRR....17....3P} are expected to have spins preferentially aligned with their orbital angular momentum. 

The exact degree of alignment and randomness predicted by both channels is still an open question. Recent cluster observations \cite{2017NatAs...1E..64C} found that the progenitor cloud's angular momentum might have a strong impact on the stellar spins, thus imprinting some preferential direction to the spins of the resulting compact objects.
For field binaries, assumptions on the supernova natal kicks, mass transfer, and tidal interactions have all been shown to be crucial in predicting the residual misalignments \cite{2000ApJ...541..319K,2008ApJ...682..474B,2013PhRvD..87j4028G,2014PhRvD..89l4025G,2016ApJ...832L...2R,2018PhRvD..97d3014W,2017arXiv170607053B}.

The first quantitative studies on inferring the formation channel of binaries using GW observations had the measurement of \emph{individual} spin parameters as their starting point. The authors of Refs.~\cite{2017CQGra..34cLT01V,2017MNRAS.471.2801S} showed that if both formation channels (dynamical and in the field) operate, their branching ratio can be measured after $\sim100$ events from the measurement of the misalignment angles.
However, a combination of the two-component spins exists which is measured better than either of them \cite{2011PhRvL.106x1101A,2010PhRvD..82f4016S,2017PhRvD..95f4053V}.%
This is the mass-weighted combination of the projection of the two spins along the orbital angular momentum, usually called the \emph{effective spin} parameter,
\begin{equation}
\chieff = \frac{\mathbf{S_1}/m_1 + \mathbf{S_2}/m_2}{m_1+m_2} \cdot \mathbf{\hat L}\,.
\label{chieffdef}
\end{equation}
Furthermore, \chieff is a constant of motion (up to at least the second post-Newtonian order~\cite{2014LRR....17....2B,2008PhRvD..78d4021R}) and it is therefore well suited to parametrize the binary evolution \cite{2015PhRvL.114h1103K,2015PhRvD..92f4016G}. 

A key signature of the formation channels is whether the intrinsic distribution of \chieff{} values is symmetric about zero.
Since formation in galactic fields is more likely to yield spins roughly aligned with the angular momentum, systems coming from this channel will have a distribution weighted towards positive \chieff (although individual events with $\chieff<0$ are possible; see Ref.~\cite{2018PhRvD..97d3014W}).
On the other hand, all spin orientations are expected to be equally likely in binaries assembled via dynamical interactions, which results in a distribution for \chieff \emph{symmetric} about zero.
The authors of Refs.~\cite{2017Natur.548..426F,2018ApJ...854L...9F} exploited this idea to show that {if all sources come from the same formation channel} the required number of BBH detections to identify it can be as small as a few tens. 
If both channels operate simultaneously, hundreds of events are required to calculate their branching ratio and characteristic parameters~\cite{2017PhRvD..96b3012T}. 
Similar reasonings have also been applied to BNS systems; see e.g.~Ref.~\cite{2017arXiv171109226Z}.

In this paper we point out several important caveats that can affect astrophysical inference based solely on the \chieff distribution of detected events.
First, the length of the GW signal depends on the sign of \chieff: for the same masses, systems with $\chieff>0$ take longer to merge and are thus easier to detect than systems with negative \chieff. 
This implies that, even if the underlying population were to be perfectly symmetric, the \chieff distribution of detected sources will show a bias toward positive values.
Second, the individual posterior distributions of \chieff present a different morphology, depending on whether the true value is positive or negative (with other binary parameters fixed). 
We show that it is easier to exclude negative values in the inferred value of \chieff if the true value of \chieff is positive, than the other way around, unless the two component masses are very different.
In addition, the uncertainty in the measurement of \chieff is systematically larger when the true value of \chieff is negative than if it is positive.
Finally, we show how these factors can bias astrophysical inference on the underlying populations. 

In the Appendixes we present an analytical toy model to explain the shape of the \chieff posteriors and a recipe to generate synthetic posteriors and likelihoods. A webpage where users can create their own realistic synthetic posteriors has been set up at \engineurl.

\section{Asymmetry in the detected \chieff distribution}\label{Sec.Detected}

If all CBCs formed via dynamical interaction, one would expect a symmetric distribution for the inferred values of \chieff, centered around zero. 
However, binaries with spins positively aligned with the orbital angular momentum have to dissipate more angular momentum and therefore take longer to merge (this is known as the orbital hang-up effect; see e.g.~Refs.~\cite{2001PhRvD..64l4013D,2006PhRvD..74d1501C,2015CQGra..32j5009S}). 
The waveform is therefore longer for systems with $\chieff > 0$ than for those with $\chieff <0$.

Given the presence of a threshold in the SNR of detectable events, this results in an observational bias in the distribution of \chieff for \emph{detectable} sources.
The net result is that even if the true population had \chieff values perfectly symmetric around zero, the \emph{detected} population will show a preference for positive \chieff. 
If not modeled, this can potentially be mistaken for the presence of a second population (e.g.~galactic field binaries) contributing preferentially to the positive \chieff branch. 

Before we report results, we should emphasize that the orbital hang-up effect is not the only reason why there might be a selection bias. Other mechanisms are known which can introduce selection effects.
For example, the template bank used by the search algorithms~\cite{2012ApJ...748..136C,2014PhRvD..89b4003P,2017PhRvD..95d2001M,2014PhRvD..90h2004D,2016CQGra..33u5004U} can introduce selection effects on various parameters.
Currently, the waveforms used for searches assume spins are aligned with the orbital angular momentum, the absence of higher-order modes, and perfectly circular orbits.
However, spin-aligned waveforms can have a large mismatch with strongly precessing waveforms at large mass-ratios. This can result on a loss of signal-to-noise ratio (SNR) and thus a selection bias that favors the observation of small or aligned spins and/or low mass-ratios~\cite{PhysRevD.94.024012}.
Likewise, sources with high eccentricity also lose SNR when matched to quasi-circular templates~\cite{2013PhRvD..87l7501H,2016PhRvD..93d3007T}, which disfavors the detection of highly eccentric binaries. In the intermediate mass regime ($M>100 M_{\odot}$), the search performance deteriorates because the astrophysical signals are shorter, and harder to distinguish from noise artifacts~\cite{PhysRevD.90.022001}.
The search performance in this regime has a strong dependence on mass and spin~\cite{PhysRevD.82.104006}.
Finally, the absence of higher-order modes in the template bank can results in a $\sim10\%$ loss in detection volume~\cite{2017PhRvD..95j4038C} in the intermediate- and extreme-mass-ratio regime.
In this work we focus on the asymmetry in the \chieff distribution introduced by the hang-up effect, and neglect other possible contributions.
To quantify the size of this effect, we simulate a perfectly symmetric distribution of sources, and we measure the distribution of \chieff of the sources which survive the SNR cut.
We simulate waveforms corresponding to the inspiral merger and ringdown of BBHs, using the {\sc IMRPhenomPv2} waveform model \cite{2015PhRvD..91b4043S,2014PhRvL.113o1101H} 

First, we draw spin directions which are isotropic on the unit sphere, such that our intrinsic population presents a symmetric \chieff distribution peaked at zero.
Next, we need to generate values for the magnitude of the individual spins. 
The impact of the orbital hang-up will critically depend on the dimensionless magnitude of the individual spins $\chi_i=|\mathbf{S_i}|/m_i\in[0,1]$.
We therefore consider five different distributions of spin magnitude $\chi$:
(i) \textit{uniform in $\chi$}: $p(\chi)=1$;
(ii) \textit{linear-low}: $p(\chi) \propto 1-\chi$;
(iii) \textit{linear-high}: $p(\chi) \propto \chi$;
(iv) \textit{Gaussian-low}: $p(\chi)=\mathcal{N}(0,0.05)$;
(v) \textit{Gaussian-high}: $p(\chi)=\mathcal{N}(0,0.25)$.
These choices facilitate comparisons with Refs.~\cite{2018ApJ...854L...9F,2017Natur.548..426F,2017PhRvL.119y1103V}.
We also include a \textit{uniform-aligned} distribution, which has uniform spin magnitude ($p(|\boldsymbol\chi \cdot \mathbf{\hat L}|)=1/2$) parallel to the orbital angular momentum ($\boldsymbol\chi \cross \mathbf{\hat L}=0$).
The component masses $m_1,m_2$ are drawn from a power-law distribution as parametrized in Ref.~\cite{2017PhRvL.118v1101A},
whereas the sky position and distance are sampled uniformly in comoving volume, with the binary randomly oriented. 

The SNR for each signal  is calculated using different noise spectral densities, representative of the expected performance improvement of the LIGO/Virgo network over the next few years: O2 (2016-17), O3 (2018-19) and Design (2020)~\cite{2016LRR....19....1A}. For O2 and O3 we use the top and bottom of the band labeled ``2016-17'' in Ref.~\cite{2016LRR....19....1A}.
Sources that have a SNR above a threshold of 8 in each detector are considered detected~\cite{2010CQGra..27q3001A,2018ApJ...861L..24L}.
We have verified that results look qualitatively similar, with a slightly smaller (larger) bias, if the more relaxed (strict) threshold of 5 (11) is used.

\begin{figure}[!ht]
\includegraphics[width=\columnwidth]{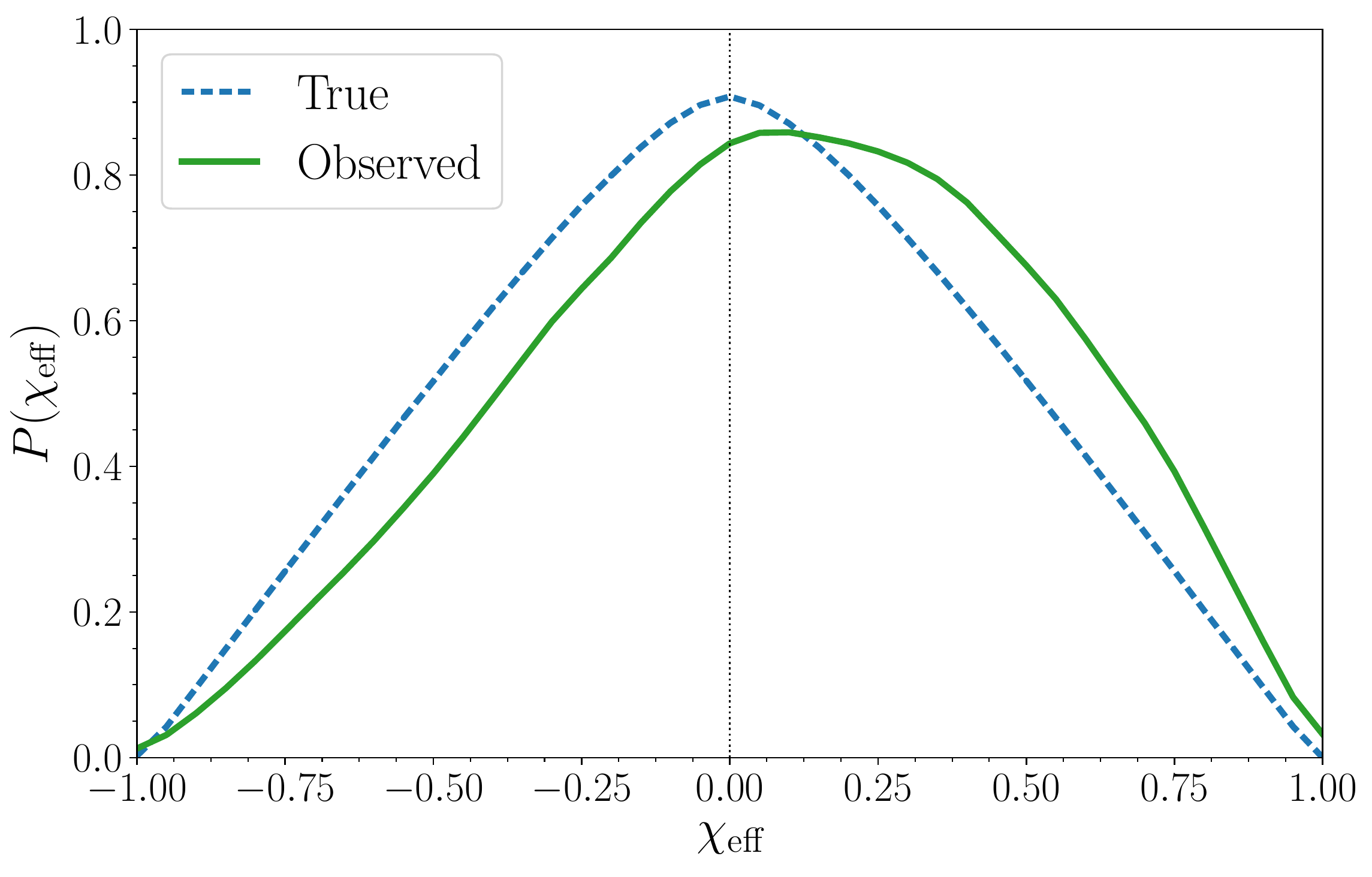}
\caption{The true (blue dashed) and observed (green solid) distribution of \chieff using the O2 sensitivity curve when the true spin magnitudes are uniform-aligned. While the true distribution is symmetric around zero, the detectable distribution shows a bias toward positive values.}
\label{Fig1}
\end{figure}

Figure~\ref{Fig1} shows the underlying (blue dashed) \chieff distribution as well as the \chieff distribution of detectable binaries (green solid) using the O2 sensitivity curve for the uniform-aligned spin distribution.
From Eq.~\eqref{chieffdef}, \chieff is defined between $-1$ (both spins are maximal and antialigned with the angular momentum) and $+1$ (both spins are maximal and coaligned with the angular momentum). Negligible values of \chieff can be due to either small spin magnitudes or spin vectors perpendicular to the angular momentum. 
The distribution of \chieff for detectable events is clearly biased toward positive values: 62\% of detectable sources have $\chieff>0$, compared to 50\% of the underlying population.
This selection bias needs to be taken into account, for example by reweighting the \chieff-dependent visible volume \cite{2018ApJ...856..173T}, or directly applying SNR selection in population models like the green curve in Fig.~\ref{Fig1}.

The uniform-aligned population shown in Fig.~\ref{Fig1}, however, corresponds to the worst-case scenario, as it has the highest probability for large values of $\chi_{\rm eff}$. 
While for this distribution $p(|\boldsymbol\chi \cdot \mathbf{\hat L}|)$ is constant, for all other distributions $p(|\boldsymbol\chi \cdot \mathbf{\hat L}|)$ decreases with $|\boldsymbol\chi \cdot \mathbf{\hat L}|$. This results in a smaller probability for \chieff to be large (in magnitude) and consequently smaller biases (Table~\ref{Tab1}). 
As expected, the bias becomes negligibly small if the population has preferentially small spin magnitudes (\emph{linear-low}). 
In the uniform-in-$\chi$ case, the effect is small enough that, even if unaccounted for, it will probably not play a role until $\mathcal{O}(100)$ of BBHs are detected (which would reduce the statistical uncertainties to the few-percent level, comparable with the bias we find).
It is worth noting that the uniform-in-$\chi$ distribution has been used as a prior in most GW data analysis to date (cf.~Ref.~\cite{2017PhRvL.119y1103V}).
For distributions with larger biases, such as the uniform-aligned one illustrated here, not only is the offset from symmetry larger, but also \emph{fewer} sources are required to obtain that level of statistical uncertainty (larger component spins are easier to measure; see e.g.~Ref.~\cite{2017PhRvD..95f4053V}).

\begin{table}
\centering
\addtolength{\tabcolsep}{+10pt}
\begin{tabular}{llll}
\hline\hline
  & O2  & O3  & Design   \\
  \hline\hline
Uniform-aligned & 12\%  & 11\%  & 11\%    \\
Uniform in $\chi$  & 6\%  & 6\%  & 6\%  \\
Linear-high & 9\% & 8\% & 7\%  \\
Linear-low & 4\% & 4\% & 4\%  \\
Gaussian-high & 3\% & 3\% & 2\%  \\
Gaussian-low & 0.3\% & 0.3\% & 0.3\% \\
\hline \hline
\end{tabular}
\addtolength{\tabcolsep}{-10pt}
\caption{The percent excess of $\chieff>0$ in the distribution of detectable binaries under various spin distributions and noise levels. $0\%$ means that the distribution is symmetric while, e.g., $+12\%$ means that 62\% of sources have positive \chieff.}\label{Tab1}
\end{table}

\section{Asymmetries in the individual posteriors} \label{Sec.Individual}
The measurement of \chieff for many, potentially hundreds, of sources is required to draw conclusions about the underlying astrophysical population. It is therefore natural to first focus on what spin inferences can be made about individual systems, and in particular how often one of the two signs of \chieff can be excluded 
(cf.~Ref.~\cite{2017PhRvD..95f4053V} for a previous partial investigation).
Thus far, a significant measurement of the sign of $\chieff$ has been possible only for GW151226, which has $\chieff>0$ \citep{2016PhRvL.116x1103A} independent of the prior \citep{2017PhRvL.119y1103V}. For GW170608, most of the posterior for \chieff is positive, but $\chieff=0$ is found within the 90\% credible interval~\citep{2017ApJ...851L..35A}. 
Posteriors for all the other events do not show a strong preference for either positive or negative values of $\chieff$. 
We argue here that parameter correlations need to be addressed carefully before strong conclusions on the underlying astrophysical population can be made.

We start by addressing the following question: if we detected a system like GW151226, but with \emph{negative} \chieff, would we be able to exclude positive values with high confidence?
To answer this question, we have created 20 simulated signals with masses and spins compatible with the estimates quoted in Ref.~\cite{2016PhRvL.116x1103A} for GW151226, rescaling the distances to achieve a SNR of either $\sim12.5$ (similar to the SNR of GW151226) or $\sim 33$ (a representative SNR for a loud source).
For each source we then create its \emph{flipped} version with the same parameters but $\mathbf{S_i} \to -\mathbf{S_i}$ (thus $\chieff \to -\chieff$) and a rescaled distance such that the SNR is unchanged. 
The last step is critical as the orbital hang-up effect would cause the system with negative \chieff to have a lower SNR, thus biasing the comparison. 
Statistical inference is then performed using the \linf{} pipeline \citep{2015PhRvD..91d2003V} and the reduced-order quadrature (ROQ) approximation to the likelihood~\citep{2016PhRvD..94d4031S}. The analyses are done with a zero-noise realization, which ensures the results are representative of the underlying physics, and not due to noise fluctuations \citep{2008PhRvD..77d2001V,2014ApJ...784..119R}.

\begin{figure}[!ht]
\includegraphics[width=\columnwidth]{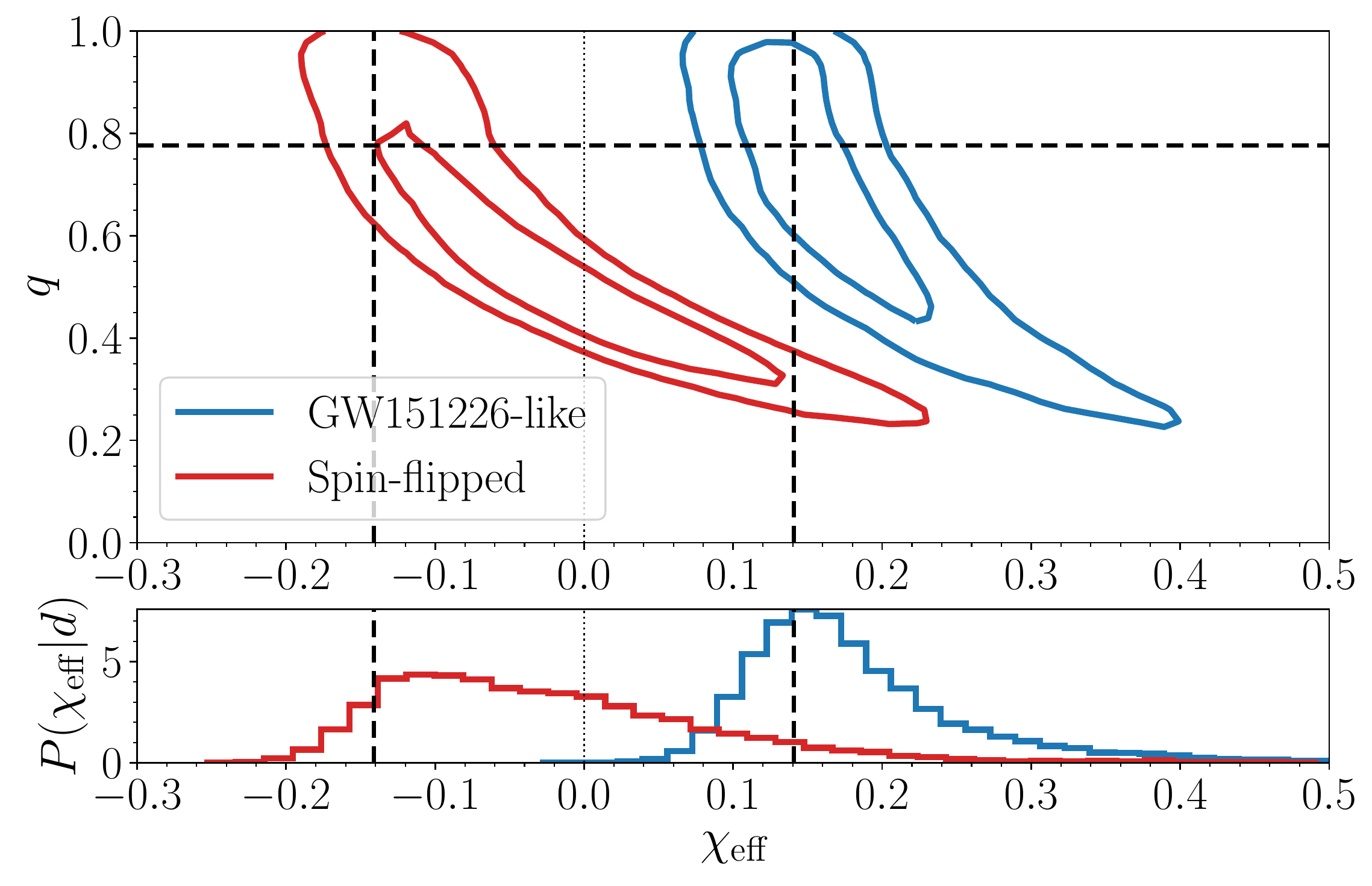}
\caption{The joint $q-\chieff$ posterior distribution for a GW151226-like BBH with positive (blue) or negative (red) \chieff. While the two events differ in the sign of \chieff, all other parameters, including the SNR, are the same. The SNR for this event is  $\sim 12.3$. The dashed lines indicate the true values of \chieff and the mass ratio.}\label{Fig2}
\end{figure}

Our results indicate that when the true \chieff is positive it is nearly always possible to exclude negative values. 
However, the opposite is not true: if the true \chieff is negative, it is rarely the case that the posterior excludes positive values. 
This is illustrated in Fig.~\ref{Fig2}, where we show the joint and the marginalized posterior distributions for \chieff and the binary mass ratio  $q=m_2/m_1 \in [0,1]$ for two of our simulated sources. 
The island on the right is the joint posterior distribution for a GW151226-like source ($\chieff\sim0.14$), while the one on the left corresponds to the same source with flipped spins ($\chieff\sim -0.14$).
For both positive and negative \chieff, the marginalized posterior distributions show a clear asymmetry around their median, with a longer tail toward larger (and positive) \chieff, and a much sharper tail on the negative side. 
The immediate consequence of this asymmetry is that, for this mass ratio, when the true value of \chieff is positive, the posterior distribution is more likely to exclude negative values than it is to exclude positive values when the true \chieff is negative.

The longer tails toward positive values of \chieff  can be understood in terms of the well-known $q-\chieff$ inspiral degeneracy~\citep{1994PhRvD..49.2658C,1995PhRvD..52..848P,2013PhRvD..87b4035B,2013PhRvD..88f4007P,2014PhRvD..89j4023C,2015ApJ...798L..17C}. 
As shown in Fig.~\ref{Fig2} below (see also, e.g., Fig.~4 in Ref.~\cite{2016PhRvX...6d1015A}), joint posterior distributions of these two parameters tend to show a pronounced degeneracy, with low (large) values of $q$ paired to large (low) values of $\chieff$.
These tails are present in all our GW151226-like simulations, regardless of the sign of \chieff. 
However, as is already visible in Fig.~\ref{Fig2}, tails tend to be more pronounced when the injected \chieff is negative. 
Ultimately, this results in higher measurement uncertainties for negative \chieff sources than positive ones. 
For our GW151226-like simulations, the 90\% credible interval for negative \chieff is typically $\sim1.5$ larger than for positive ones.
We discuss a simple model for these features further in Appendix \ref{Sec.Toy}.

It is worth stressing that this degeneracy is only present in the inspiral part of the signal.
It is therefore expected to be milder for heavier BBHs, which present fewer inspiral cycles in band, and more  prominent for lighter BBHs, such as GW151226, and BNSs.
In order to verify this, and to check the generality of these trends, we generate another set of simulated signals covering a broader range of system parameters. 
We consider BBHs with spins  $0\leq |\chieff|\leq 0.7$ (in steps of 0.1 in \chieff) and seven different values of detector-frame component masses, $(m_1,m_2)=$ (30-30), (30-15), (30-10), (15-15), (15-7.5), (15-5), (5-5)~\msun. 
We also simulate BNSs with $(m_1,m_2)=$ (2.2-1.3), (2.0-1.4), (1.4-1.4)~\msun and effective spins $0\leq |\chieff|\leq 0.2$ (in steps of $0.05$). 
We generate the BNS signals with  the same waveform family used for BBHs \citep{2015PhRvD..91b4043S,2014PhRvL.113o1101H} and do not include tidal terms.
For both BNSs and BBHs, we considered two values of spin tilts ($10^\circ$ and $30^\circ$), which are defined as the angle between the spins and the angular momentum vector at a GW frequency of 20~Hz, and two values of the network SNR (15 and 30).

\begin{figure}[!ht]
\includegraphics[width=\columnwidth]{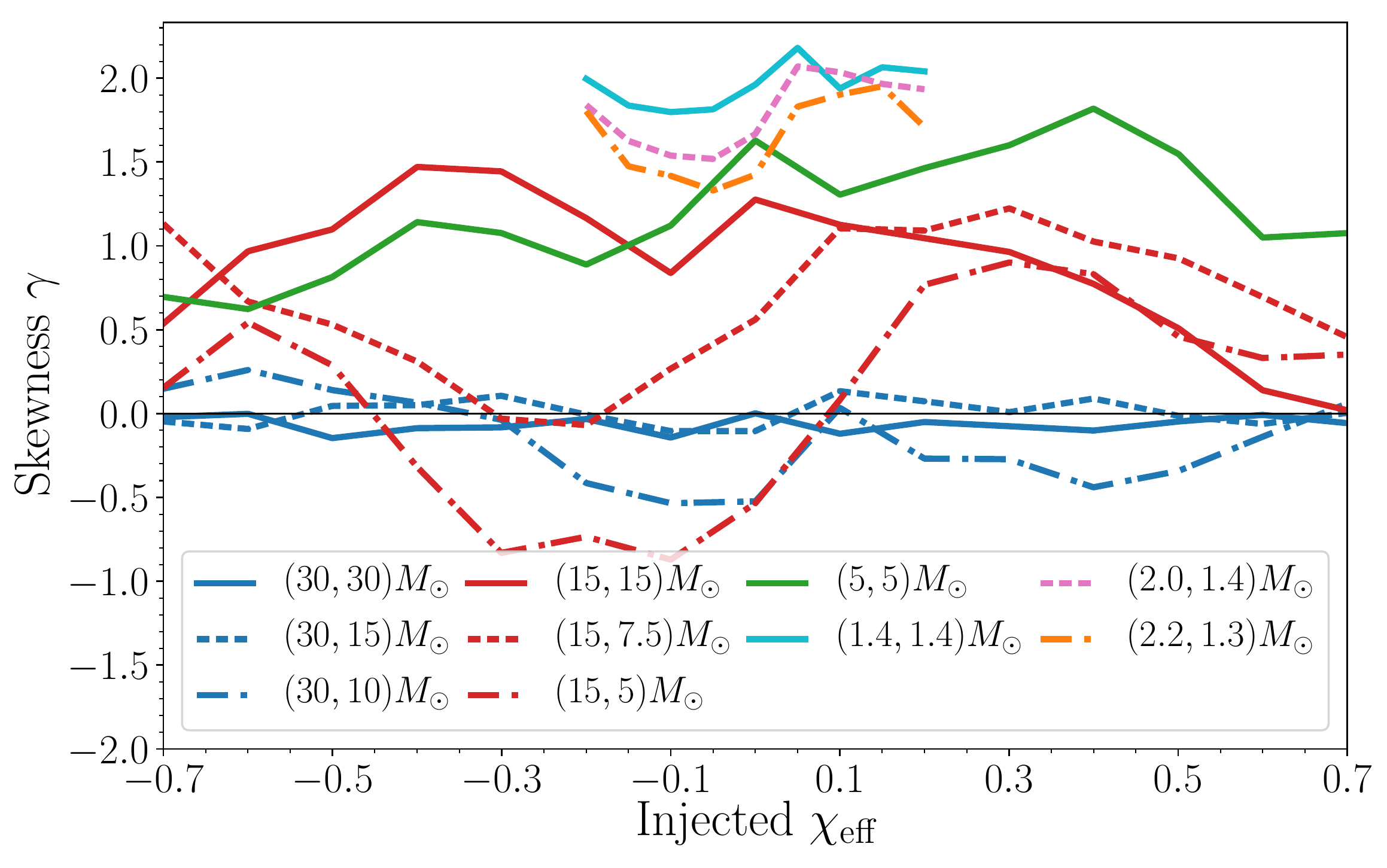}
\caption{The skewness coefficient for the \chieff posterior distributions of all simulated runs. Component masses are given in the legend. For most runs, the skewness is positive, indicating a large tail toward larger \chieff.}\label{FigSkew}
\end{figure}

Figure~\ref{FigSkew} shows the skewness coefficient $\gamma$ of the marginalized \chieff posterior for injections with different masses at a fixed SNR of 30 and spin tilt angle of $10^\circ$ (other runs present very similar results, cf.~Appendix~\ref{Sec.StatTables}). 
Here the skewness coefficient $\gamma$ is also known as the third standardized moment of a probability distribution,
\begin{align}
\gamma = \mathbb{E}\left[ \left( \frac{X-\mu}{\sigma} \right)^3 \right],
\end{align}
where $\mathbb{E} \left[ \cdot \right]$ is the expectation operator, $X$ is the random variable, $\mu$ is the mean and $\sigma$ is the standard deviation.
Values of $\gamma>0$ indicate a posterior distribution with a larger tail towards positive values.
The skewness is positive for all $q=1$ and $q=1/2$ BBHs, and for all BNSs.

The skewness approaches zero for heavy BBHs, which is expected since the merger and ringdown phase breaks the $q-\chieff$ degeneracy.
Conversely, the skewness increases for BNS injections, reaching $\gamma\sim2$. For comparison, the skewness for the two posteriors in Fig.~\ref{Fig2} are $\gamma=1.0$ (1.4) for positive (negative) \chieff.

A few of our simulated sources with $q=1/3$ present negative values of $\gamma$. These are cases where \chieff and $q$ are generally \emph{not} measured well, with either a posterior peak not centered at the true value or a very broad posterior distribution (or both); see Appendix~\ref{Sec.StatTables}.
For these systems with small $q$, the best measured parameter is a different combination of masses and spins, namely the 
1.5 post-Newtonian (PN) phase coefficient~\citep{1992PhRvD..46.1517W,1993PhRvD..47.4183K,1994PhRvD..49.2658C,1995PhRvD..52..848P}.
We verified that for all sources with negative skewness the 1.5PN coefficient is measured accurately, although \chieff is not. 
Furthermore, we have verified that in our recovered posteriors this phase term is approximately uncorrelated with the mass ratio.

In Appendix~\ref{Sec.Toy} we present a simple toy model that shows how Gaussian and uncorrelated likelihoods for the  mass ratio and the 1.5PN phase term can result in a skewed likelihood and posterior for \chieff. 
By generating Gaussian likelihood distributions for the symmetric mass ratio and the 1.5PN phase term, performing the appropriate change of coordinates, and taking into account prior bounds, we find that the $q$--$\chieff$ likelihood profiles have shapes similar to what is obtained from the actual posterior samples and shown in Fig.~\ref{Fig2}.

\begin{figure}[!ht]
\includegraphics[width=\columnwidth]{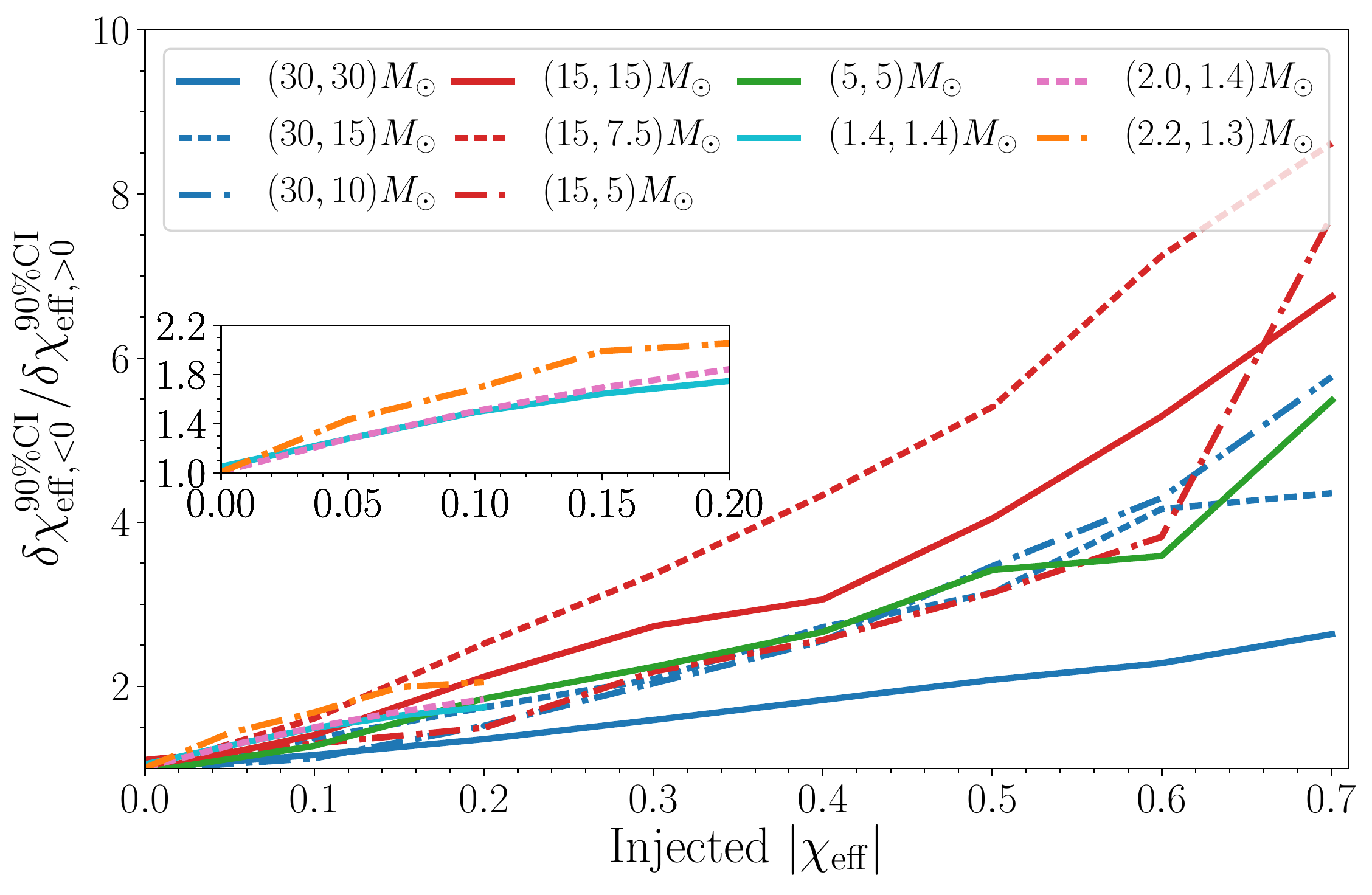}
\caption{Ratio between the 90\% credible intervals for sources with negative over positive \chieff. The $x$~axis gives the absolute value of the simulated \chieff. Component masses are given in the legend. %
}\label{Fig4}
\end{figure}

Returning to our discussion of the injected and recovered signals, we find that irrespectively of the sign of $\gamma$, negative \chieff values are harder to constrain than positive ones.
Figure~\ref{Fig4} shows the ratios of the 90\% credible intervals for \chieff of our systems and of their spin-flipped versions.
As expected, we find that this ratio depends on the total mass, and it is smaller for heavier systems. For 30-30~\msun BBHs the ratio reaches at most $\sim2$.
Lighter systems present much larger ratios. Sources of 15-7.5~\msun, similar to GW151226, can have an uncertainty for negative \chieff$=-0.4$  up to 4 times larger than for $\chieff=+0.4$. This factor reaches $\sim 8$ for $|\chieff|\sim0.7$. 
The \chieff$\sim 0.1$ case is similar to the GW151226-like simulations presented above, where the uncertainty in the measurement of negative \chieff is typically 1.5 times larger than that of the same event with positive \chieff. 
BNS sources follow a similar trend, with ratios of $\sim 2$ at $\chieff=|0.2|$. Already at $|\chieff|=0.05$, uncertainties for negative \chieff can be a few tens of percent larger than for positive ones. 
The ratio of standard deviation looks numerically similar.

We conclude this section with a discussion on the accuracy of the individual \chieff posteriors and the role of priors.
As shown in Ref.~\cite{2017PhRvL.119y1103V} the measurement of \chieff can be significantly impacted by the priors one is using in the Bayesian analysis.
In the analyses presented in this section, we have used the same priors used by the LIGO-Virgo Collaboration for all the events detected up to the second science run (the authors of Ref.~\cite{2017PhRvL.119y1103V} referred to these priors as P1). 
For each compact object, the prior of the spin direction is random on the unit sphere, while the prior of the spin magnitude is uniform in the range $[0,0.89]$, and is discussed further in Appendix~\ref{Sec.Toy}.
This results in a prior for \chieff which is peaked at zero (cf.~Fig.~5 of Ref.~\cite{2017PhRvL.118v1101A}).

\begin{figure}[!ht]
\includegraphics[width=\columnwidth]{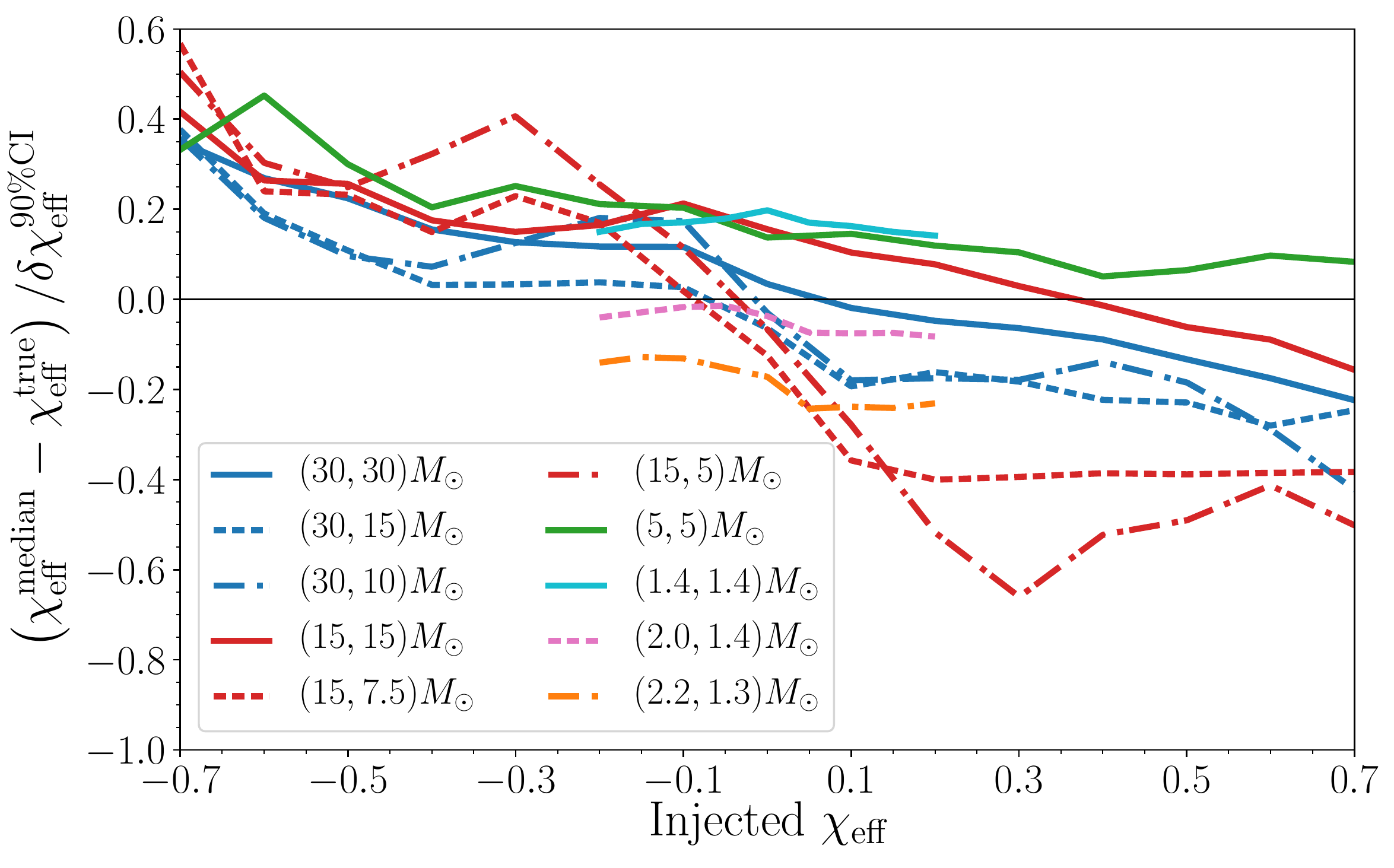}
\caption{The offset in the median of the \chieff posterior for the simulated events with a SNR of 30 and tilt (for the BHs only) of $10^\circ$, normalized by the width of the 90\% credible interval. See Appendix~\ref{Sec.StatTables} for medians of more choices of SNRs and spins.}\label{FigOffset}
\end{figure}

We find that this choice leaves a clear imprint in the posterior distribution of the individual events, pushing their medians toward the region of higher prior, i.e. toward zero. This is shown in Fig.~\ref{FigOffset}, where on the $y$~axis we report the difference between the median \chieff posterior and the true value, divided by the 90\% uncertainty for the systems with SNR 30 and BBH tilt of $10^\circ$.
For all the events, the shift is indeed in the direction of \chieff closer to zero, and usually smaller than half of the 90\% credible interval.
The three curves for the BNS sources have been shifted vertically as described in Appendix~\ref{Sec.ROQ}.
For the runs with a SNR of 15, the shifts are larger, consistent with the intuition that the prior should matter more for weaker signals. These findings are in agreement with the results of Ref.~\cite{2018arXiv180403704C} on the difficulty of measuring large spins from GW data.
We tabulate all relevant statistics in Tables~\ref{Tab.CIBBH_SNR15_tilt10}--\ref{Tab.BiasBNS_SNR30_tilt0}.

\section{Impact on astrophysical inference}\label{Sec.Astro}

All the effects we have described thus far can, in principle, affect how precisely and accurately one can study the spin distribution of the detected GW sources, and infer their formation channels. In Sec.~\ref{Sec.Detected} we saw how the population of the detectable BBH sources will have higher \chieff than that of the true underlying population. In Sec.~\ref{Sec.Individual} we have shown that negative \chieff are harder to measure, and that the posterior distributions for \chieff are often skewed and affected by the priors.

It is natural to ask which of these factors can significantly affect astrophysical inference. For example, the effects of priors should not matter when one does hierarchical modeling, since in that case the likelihood is used, that is: one divides the posterior by the prior used in the analysis.
On the other hand, the correlation between mass ratio and \chieff can impact inference solely based on \chieff.

As an example, we repeat one of the analyses performed in Ref.~\cite{2017Natur.548..426F}. In particular, we assume that two formation channels are possible: one which results in an isotropic distribution of spins, and the another which gives roughly aligned spins.
Here we focus on the effect of the correlation and do not include selection bias in our simulated population; as discussed in Sec.~\ref{Sec.Detected} the selection bias can be readily removed.
We assume that all black hole sources come from a channel that results in uniform spin magnitudes and isotropic orientations (this is the model ``flat-isotropic'' or FI in Ref.~\cite{2017Natur.548..426F}). 
We focus on 5-5~\msun BBH sources, and assume that all of them have a SNR of 15.
We create a catalog of synthetic BBH sources, having true \chieff drawn from the isotropic distribution (for simplicity, we neglect at first the selection bias described in Sec.~\ref{Sec.Detected}; its effect will be described later),
and for each of them we create a synthetic posterior distribution. Our goal is to calculate the odds ratio between the isotropic spin and the aligned-spin models (this is the model ``flat-aligned'' or FA in Ref.~\cite{2017Natur.548..426F}): $\mathcal{O}^\mathrm{align}_\mathrm{iso}$.

\begin{figure}[!ht]
\includegraphics[width=\columnwidth]{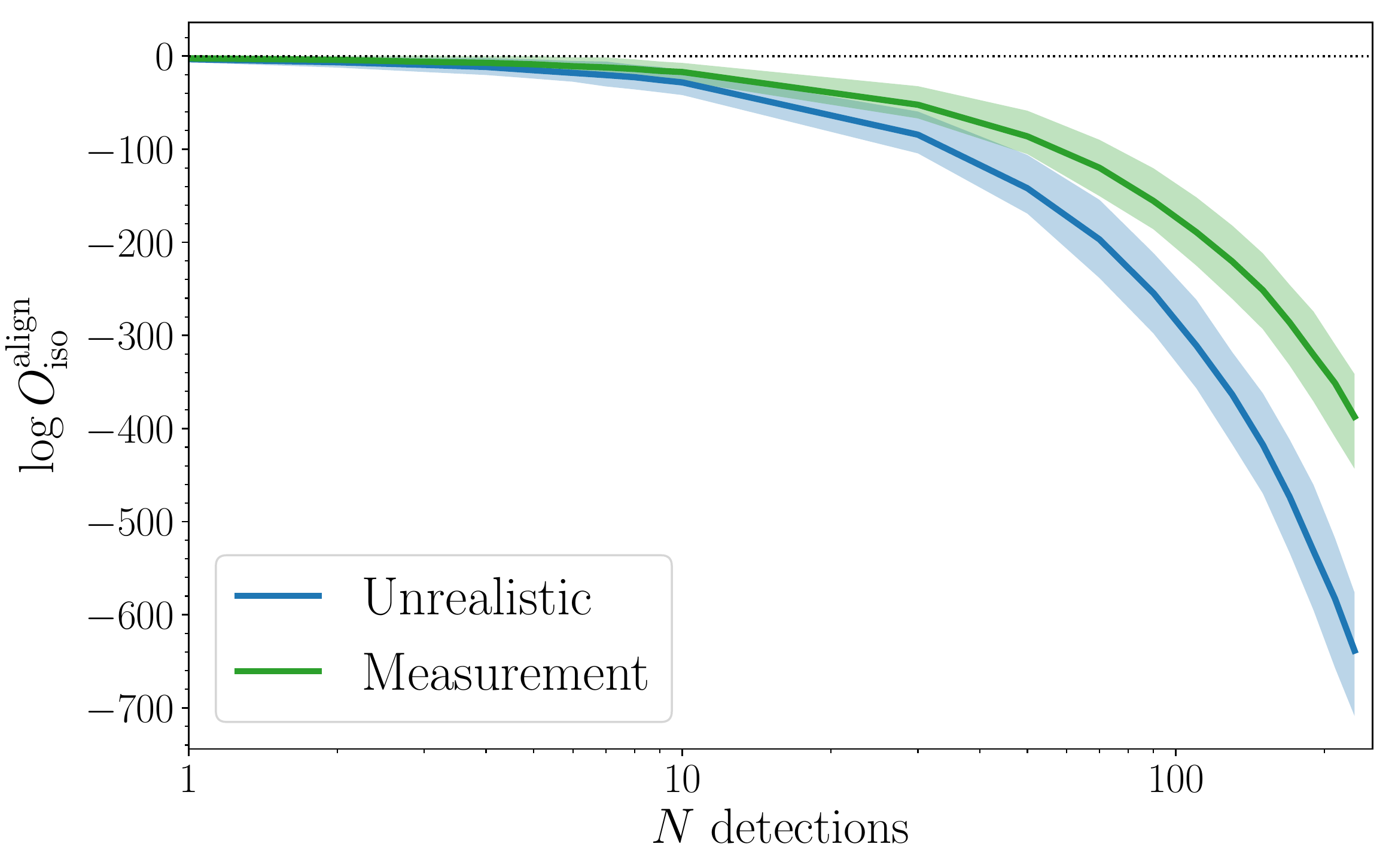}
\caption{Odds ratio between a model where all events have aligned spins vs one where spins are isotropic. The true population has isotropic spins. This is for 5-5~\msun and a SNR of 15. The unrealistic curve assumes that all \chieff posteriors are centered at the true value. }\label{Fig.Odds}
\end{figure}

First, we perform an (unrealistic) analysis where each \chieff posterior is a perfect Gaussian centered at the true value with a width of $0.1$. Each posterior is divided by its \chieff prior to obtain likelihood distributions suitable for a hierarchical analysis. Using this approach we obtain the curve labeled ``Unrealistic'' in Fig.~\ref{Fig.Odds}.
Then, we repeat the same analysis by generating synthetic posterior distributions for all sources using the recipe we provide in Appendix~\ref{Sec.Recipe}, which results in skewed posteriors. 
Critically, we do \emph{not} assume that the posteriors are centered at the true values, but rather we use the results from the previous section to inform the typical offset. 
Here too we divide each posterior by its prior. 
These results are labeled ``Measurement" in Fig.~\ref{Fig.Odds}.

Figure~\ref{Fig.Odds} shows how the cumulative natural log of  $\mathcal{O}^\mathrm{ali}_\mathrm{iso}$ evolves as a function of the number of detected sources.
The error band reports the spread on the measurement obtained by creating 100 random realizations of the catalog.

We see that the odds ratios for the generalized model are less negative (i.e.~favor the isotropic model less) than what is obtained with the unrealistic Gaussian model centered at the true values.
This is consistent with the fact that for light equal-mass systems posteriors and likelihoods are typically biased toward higher \chieff (see Tables~\ref{Tab.BiasBBH_SNR15_tilt10}, \ref{Tab.BiasBBH_SNR30_tilt10}, \ref{Tab.BiasBBH_SNR15_tilt30} and \ref{Tab.BiasBBH_SNR30_tilt30}).
After less than 100 events, the results obtained with a realistic approach vs one in which the true positions of \chieff are known start being clearly different.

The fact that in this case the generalized model does ``worse'' (gives odds which favor the right model less strongly) is just a consequence of having simulated a population with only perfectly isotropic spins.
In a more general situation, both approaches would give biased answers, for different reasons.

\begin{figure}[!ht]
\includegraphics[width=\columnwidth]{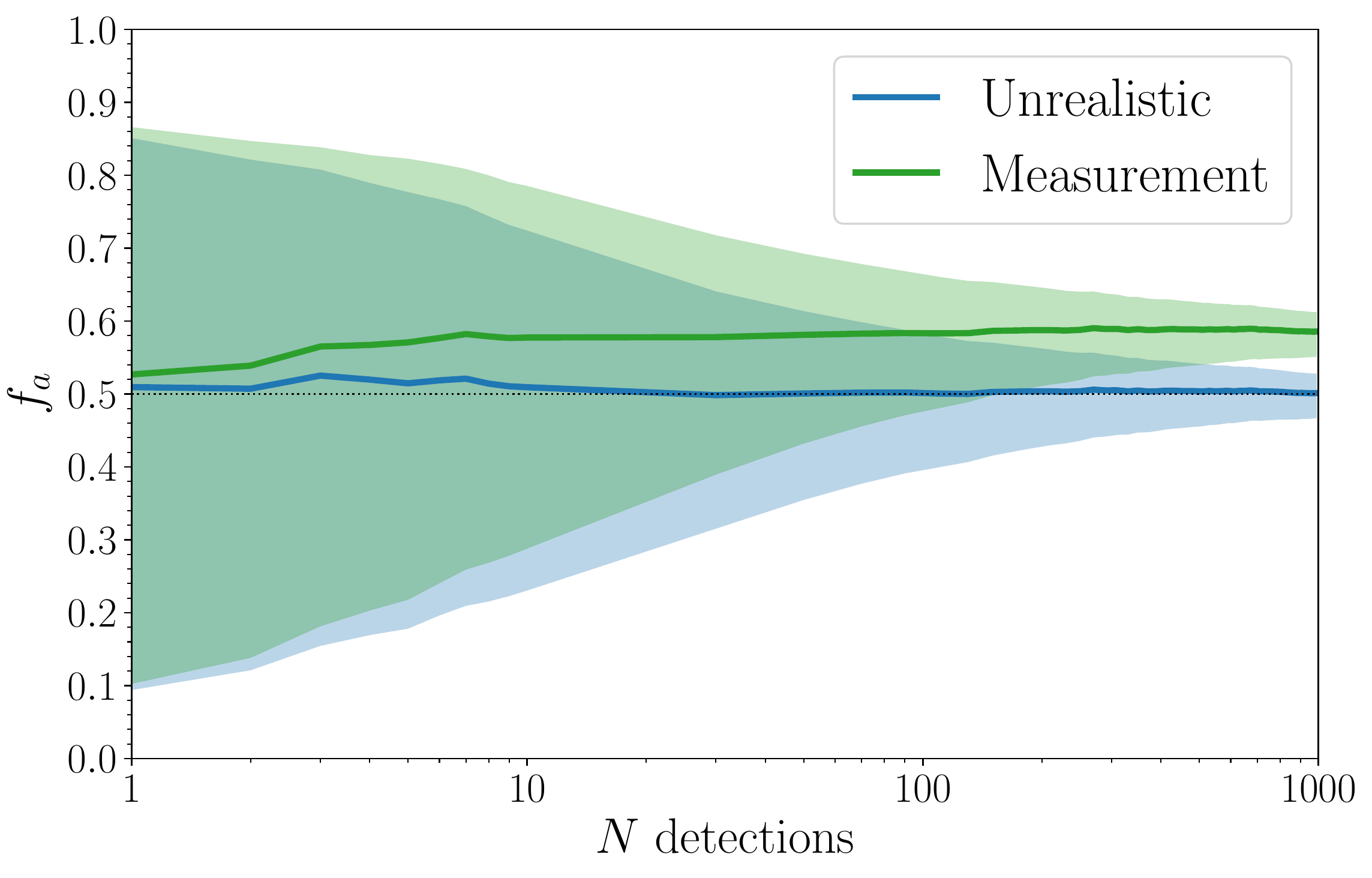}
\caption{The posterior on the fraction of aligned-spin sources, \fa, as a function of the number of detected events. This is for 5-5~\msun and a SNR of 15. The unrealistic curve assumes that all \chieff posteriors are centered at the true value.}\label{Fig.Pop}
\end{figure}

To see this we create catalogs where both aligned and isotropic spins are present. Specifically, a fraction \fa of events will have aligned spins drawn from the flat-aligned distribution, while a fraction $(1-\fa)$ will have random spins, coming from the flat-isotropic distribution. 
First, we perform inference using the realistic posteriors informed by our Markov chain Monte Carlo runs, this is the curve labeled ``Measurement'' in Fig.~\ref{Fig.Pop}, where we show the posterior distribution for \fa as a function of the number of detected events, when the true underlying value of \fa is 0.5. 
For each value of $N$, we create 100 random catalogs of $N$ BBHs and calculate the posterior distribution of~\fa.
Here, too, we focus on 5-5~\msun BBH sources, and assume that all of them have a SNR of 15.
The colored bands in Fig.~\ref{Fig.Pop} are the 90\% credible intervals averaged over the 200 catalogs.
We see how a clear bias is present: after roughly 180 events, the true value of \fa is excluded with high confidence. Ultimately, the measurement converges to a value of $\fa\sim 0.58$.
To verify that the algorithm works as expected, we also consider the (unrealistic) case where each posterior distribution is a Gaussian centered at the true value, with a width informed by our MCMC simulations. 
This is labeled ``Unrealistic'' in Fig.~\ref{Fig.Pop}, and we see how in this idealized scenario the posterior for \fa converges to the true value.

We have verified that the main contributor to the bias in the ``Measurement'' curve is not the skewness of the individual posteriors, but rather the fact that for 5-5 \msun BBHs, the median of the \chieff posterior is systematically offset to the right in our simulations.
This median-bias depends quite strongly on the actual properties of the underlaying distribution, including mass and mass ratio.
For example, if most of the detected BBHs were heavy and roughly equal mass, more detections would be required for the bias to be significant.
Our analysis suggests that while tests based on a single parameter, \chieff in this case, might yield reasonable results when only a few tens of sources are detected, in the long term more sophisticated methods will be required. To properly account for correlations and selection effects, higher-dimension hierarchical models should be considered, where all relevant parameters and hyper-parameters are measured at once. 
This, in turn, might increase the number of sources required to achieve a given level of precision.

\section{Conclusions} 

Measurements of the effective spin \chieff with GW observations can shed light on the formation pathways of black hole and neutron star binaries. In particular, a population with a \chieff distribution symmetric about zero is generally believed to be a solid signature of dynamical formation channels.

In this paper we have shown that such astrophysical statements must be made with care. 
Even if the underlying population were perfectly symmetric in \chieff, several kinds of asymmetries and selection biases affect the final measured distribution.

Because of the orbital hang-up effect, systems with positive \chieff have longer inspirals and hence higher SNRs. They will therefore be detected more easily than sources with negative \chieff. Significant asymmetries also exist in the analysis of \emph{each} event. Due to correlations between the mass ratio and the effective spin, the posterior distribution of \chieff will typically present a prominent tail toward larger (and more positive) \chieff, while no significant tail is present toward more negative values. Furthermore, these tails are generally fatter for  sources with negative true \chieff.

This observation carries two key consequences: (i) excluding negative \chieff when the true value is positive is more likely than excluding positive values when the true \chieff is negative, and 
(ii) if the true \chieff is negative, measurements of it come with larger uncertainties, at fixed SNR.

In other words, measuring positive \chieff is easier, as tentatively confirmed by the GW detections reported to date.
For light BBHs like GW151226, we find that an injected negative \chieff yields 90\% credible intervals 150\% larger than an identical system with a positive \chieff of the same magnitude and same SNR. 
These effects are milder for heavier BBHs, as they accumulate significant signal to noise from the merger and ringdown phases, which helps break the degeneracy between \chieff and mass ratio. Conversely, BNSs suffer even more from these observational effects. 

It is worth noting that most of the existing studies in the literature that use \chieff to infer properties of the underlying population have generated synthetic Gaussian posteriors distributions for \chieff, centered at the true values, with uncertainties informed by the LIGO's detections~\citep{2018ApJ...854L...9F,2017Natur.548..426F,2017arXiv171109226Z}. 
These studies do not differentiate in any way the correlations present in sources with opposite signs of \chieff and thus do not capture the different morphologies that we have documented in this work. 

We have provided a simple recipe to produce synthetic \chieff distributions that are more representative of what is encountered in the actual analysis of gravitational-wave data and used them to verify how the effects described in this paper affect astrophysical inference.
We have shown that when using more realistic posterior distributions for \chieff, astrophysical inferences may be biased. However, this will not be a problem until a few hundreds of detections are made. 
Once hundreds of events are available, one should use a more elaborate inference scheme, in which all relevant parameters are measured at once.
While methods have been proposed to account for low-significance or undetectable sources~\cite{2013NJPh...15e3027M}, this is still an area of very active research~\cite{2018arXiv180902063M,2018arXiv180506442W,2018arXiv180608365T}.
Multidimensional inference is bound to increase the overall uncertainty, implying that uncovering the formation pathways of compact binaries is likely to require more events than claimed in existing work.

\acknowledgments

K.N. and S.V.~acknowledge support of the MIT physics department through the Solomon Buchsbaum Research Fund, the National Science Foundation, and the LIGO Laboratory. 
LIGO was constructed by the California Institute of Technology and Massachusetts Institute of Technology with funding from the National Science Foundation and operates under cooperative agreement PHY-0757058.
D.G. is supported by NASA through Einstein Postdoctoral Fellowship Grant No.~PF6--170152 by the Chandra X-ray Center, operated by the Smithsonian Astrophysical Observatory for NASA under Contract NAS8--03060.
The authors acknowledge the LIGO Data Grid clusters.
LIGO Document Number P1800103.

\appendix
\section{Model for skewed posteriors}\label{Sec.Toy}

In this appendix, we present a simple model which recovers the skewed posterior probabilities for \chieff.
We consider a PN approximation to the frequency domain waveform, truncating at the 1.5PN term where the first spin contributions appear~\citep{1993PhRvD..47.4183K,2014LRR....17....2B}.
The 1.5 PN contribution to the phase depends on four parameters: the chirp mass $\mathcal M$, the symmetric mass ratio $\eta=q/(1+q)^2$, and the components of each of the two spin vectors aligned with the orbital angular momentum, $\chi_{z,i}$.
We can alternatively parametrize the spin dependence by a different pair of independent spin variables, for example \chieff and $\chi_a = (\chi_{z,1} - \chi_{z,2})/2$.
Using the stationary phase approximation, the 1.5PN phase term can be written as \citep{1994PhRvD..49.2658C} 
\begin{align}
\label{eq:psi1.5}
\psi_{1.5} &= (\pi \mathcal M f)^{-2/3} \psi  \,, 
\end{align}
where
\begin{align}
\label{eq:psi}
\psi & =   \eta^{-3/5}\left[  \frac{(113 - 76 \eta) \chieff + 76 \, \delta \, \eta \,  \chi_a}{128} - \frac{3\pi}{8}\right] \,,
\\
\label{eq:delta}
\delta & = \frac{m_1 -m_2}{m_1+m_2} \,.
\end{align}
\begin{figure}[tb]
\includegraphics[width=0.98\columnwidth]{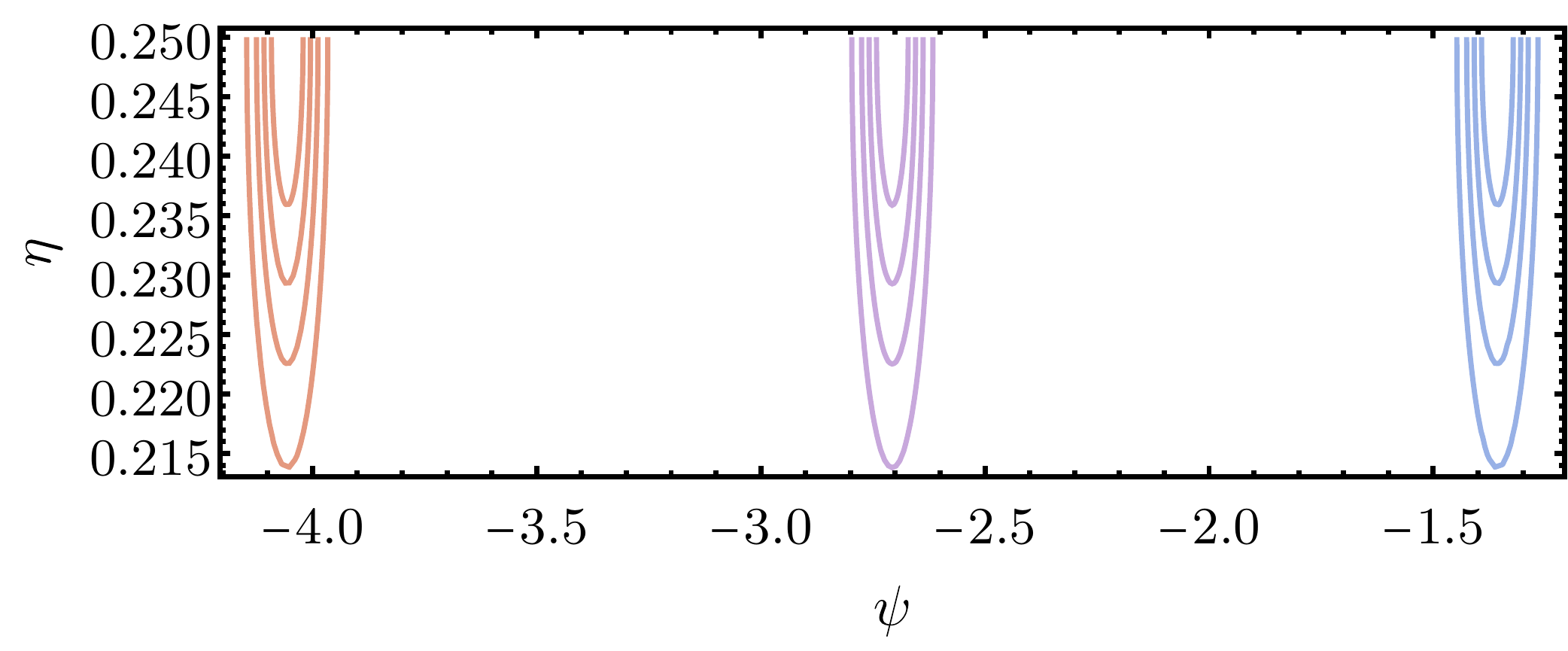} \\
\includegraphics[width=0.98\columnwidth]{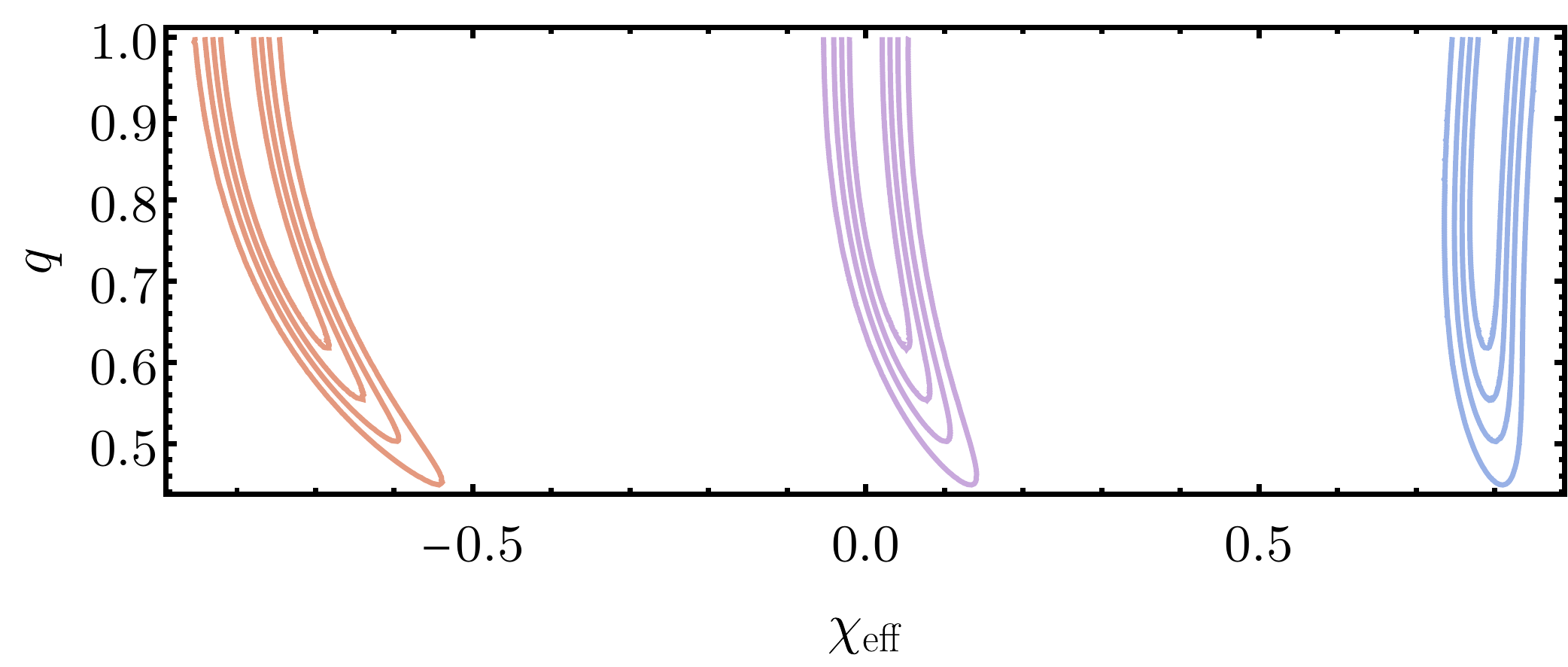} \\
\includegraphics[width=0.98\columnwidth]{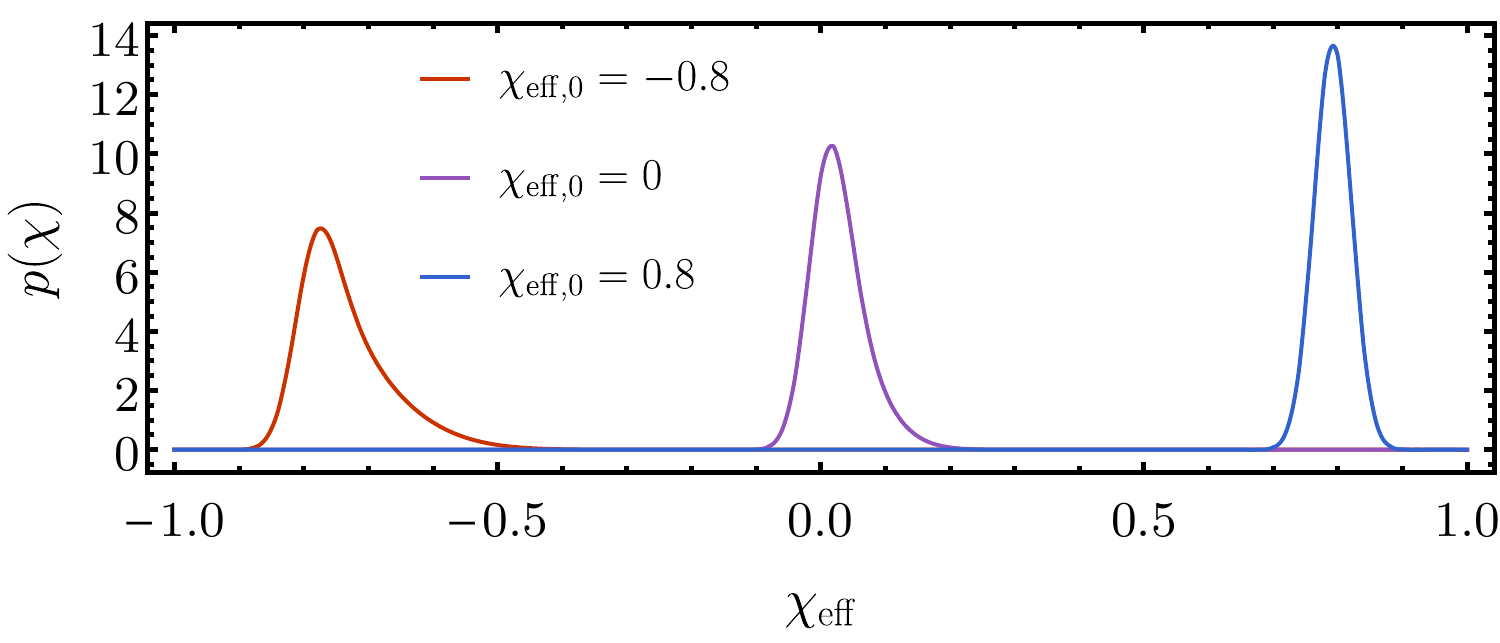}
\caption{Model for the skewness of \chieff posteriors, using Gaussian likelihoods for $\eta$ and $\psi$ and flat priors in $q$ and $\chieff$. The likelihoods are centered on $(\eta_0,\chi_{\rm eff,0}) = (0.25, -0.8)$ (red), $(\eta_0, \chi_{{\rm eff},0}) = (0.25, 0)$ (purple) and $(\eta_0, \chi_{{\rm eff},0}) = (0.25, 0.8)$, (blue), with $\sigma_\eta = 0.02$ and $\sigma_\psi = 0.05$. {\it Top:} Equally spaced contours of constant likelihood in $(\eta, \psi)$ coordinates. {\it Middle:} Equally spaced contours of constant posterior probability in $(q,\chieff)$ coordinates, using flat priors in $q$ and $\chieff$. {\it Bottom:} Marginalized \chieff posteriors, displaying positive skew and shifted maxima.}
\label{fig:ModelPos}
\end{figure}
The coefficient of this phase term is measured accurately from the GW signal, and we have verified that in our simulated and recovered signals it is approximately uncorrelated with the mass ratio.
It is the coordinate transformation from $\psi$ to \chieff, along with the physical requirement that $\eta \leq 0.25$, that allows to explain the shape of the two-dimensional $q-\chieff$ posteriors and the skewness of the marginalized \chieff posteriors. 

To see this, we consider an uncorrelated Gaussian likelihood in $\eta$ and $\psi$ for measured GW strain data $d$.
In this model, we fix $\mathcal M$ to a fiducial value in order to represent the fact that $\mathcal M$ is well measured.
As a further simplification, we neglect the spin on the less massive body, setting $\chi_{z,2} =0$, so that $\chieff = \chi_{1,z}/(1+q)$ and $\chi_a = \chi_{1,z}/2$. 
This results in a two-dimensional model, and the likelihood factorizes as
\begin{align}
\mathcal L(d|\eta, \psi) = \mathcal N(\eta_0, \sigma_\eta) \mathcal N(\psi_0, \sigma_\psi) \,.
\end{align}
Importantly, in this model it is possible for the likelihood to extend to the unphysical regime $\eta > 0.25$ and $|\chieff| > 1$, and it is the priors that restrict $\eta$ and $\chieff$ to their physical ranges.
The top panel of Fig.~\ref{fig:ModelPos} illustrates three example likelihoods selected to mimic the recovered posteriors for our injected signals.
Next, we derive the resulting posterior probability densities after changing parameters from $\theta^a = (\eta, \psi)$ to $\theta^{a'} = (\eta,\chieff)$. 
Recall that the likelihood of the data given a GW model and set of extrinsic parameters is the same whether we label those extrinsic parameters with $\theta^{a'}$ or $\theta^a$.
Thus the likelihood simply transforms as $\mathcal L(d|q,\chi) = \mathcal L(d|\eta(q),\psi(q,\chieff))$.

Meanwhile, the posterior probability is proportional to the product of the likelihood and the prior probabilities on the extrinsic parameters  $p(\theta^a)$. 
These priors transform according to the Jacobian of $\partial \theta^{a'}/\partial \theta^a$, but we can simply state the priors in terms of $q$ and $\chieff$ directly.
To keep our analysis simple, we select flat priors in $0 \leq q \leq 1$ and $-1 \leq \chieff \leq 1$, so that in these ranges our model for the posterior is
\begin{align}
p(\theta^{a'}|d) \propto p(q,\chieff)\mathcal L(d|q,\chieff)  \propto \mathcal L(d|q,\chieff) \,.
\end{align}
We plot the three posteriors in $(q,\chieff)$ in the middle panel of Fig.~\ref{fig:ModelPos} for the same three likelihoods illustrated in the top panel.
The resulting posteriors takes on the characteristic ``banana'' shape seen in the posteriors recovered from our simulated signals.

Finally we can consider the marginalized posteriors in $\chieff$ in this model. 
We find that these posteriors have positive skew, as seen in Fig.~\ref{fig:ModelPos}. 
It is clear from the likelihoods how the tilted posteriors and boundary of $\eta$ generate skewed posteriors once projected onto \chieff, and why the effect is greater for likelihoods peaked at more negative \chieff.
In addition, the maximum of the posterior is shifted rightward relative to the maximum likelihood in all cases.

We note that while previous studies have discussed how the combination \citep{1994PhRvD..49.2658C,1995PhRvD..52..848P}
\begin{align}
\beta = \frac{113-76\eta}{12} \chieff + \frac{76 \,\delta\, \eta}{12} \chi_a
\end{align} 
is better measured than $\chieff{}$ during the inspiral and reduces degeneracies (see e.g.~Ref.~\cite{2013PhRvD..88f4007P}), here we find that it is $\psi$ that is well measured and weakly correlated with $\eta$. We also find that it is the mapping between $(\eta, \psi)$ and $(q,\chieff)$ which reproduces the observed degeneracy.

\section{A recipe for generating simulated \chieff posteriors}\label{Sec.Recipe}

In this appendix we provide a simple recipe for generating simulated posterior distributions for \chieff which go beyond the simple Gaussian approximation, and include the effects we have described in Sec.~\ref{Sec.Individual}.

We find that posterior distributions for \chieff can be parametrized well with a generalized normal distribution (GND) of type II~\citep{hosking_wallis_1997}.

The GND of a random variable $x$ can be parametrized by a scale $\alpha$, a location $\xi$ and a shape $\kappa$, as
\begin{align}\label{eq:GND}
p(x)=\frac{\phi(y)}{\alpha-\kappa(x-\xi)},
\end{align}
where $\phi(y)$ is the standard normal distribution of a random variable $y$, defined as
\begin{align}\label{eq:why}
y \left( x;\alpha,\xi,\kappa \right) =
\begin{cases}
- \frac{1}{\kappa} \log \left[ 1- \frac{\kappa(x-\xi)}{\alpha} \right] & \text{if } \kappa \neq 0, \\ \frac{x-\xi}{\alpha} & \text{if } \kappa=0.
\end{cases}
\end{align}

To generate a synthetic posterior distribution for \chieff, we can relate the median $\tilde{x}$, standard deviation $\sigma$ and skewness $\gamma$ of the \chieff posterior to GND parameters as follows:
\begin{align}
\tilde{x}&=\xi, \label{eq:median} \\
\sigma^2&=\frac{\alpha^2}{\kappa^2} e^{\kappa^2} \left( e^{\kappa^2} - 1 \right), \label{eq:variance} \\
\gamma&=\frac{3 e^{\kappa^2} - e^{3 \kappa^2} - 2}{(e^{\kappa^2} - 1)^{3/2}} \text{ sign}(\kappa) \label{eq:skewness}.
\end{align}
These can be inverted to obtain the $\alpha$, $\xi$ and  $\kappa$ necessary to simulate the posterior.

\begin{figure}[!ht]
\includegraphics[width=\columnwidth]{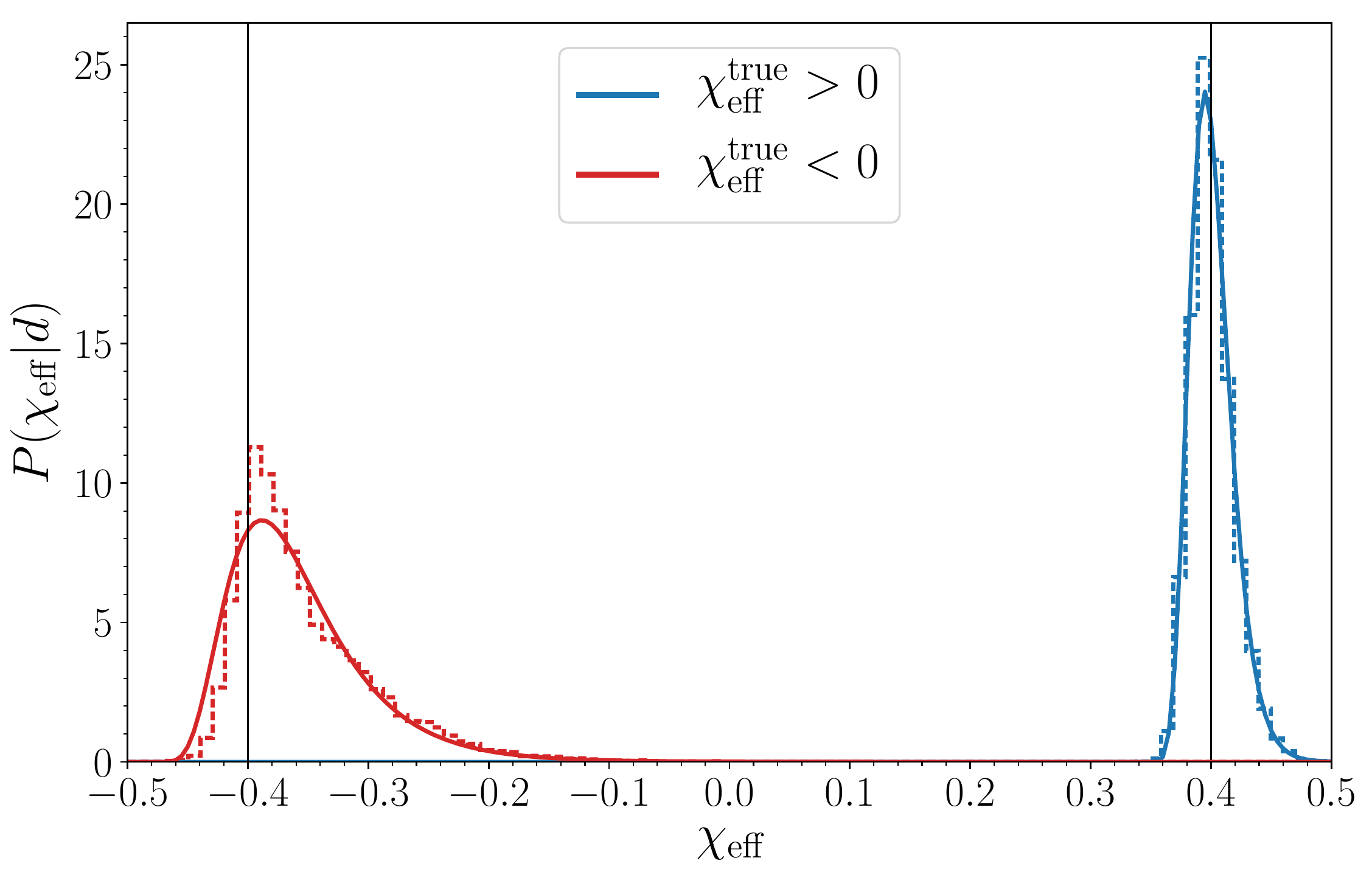}
\caption{Posterior distribution for \chieff for two example BBH sources (histograms) together with the synthetic posteriors produced using the method described in Appendix~\ref{Sec.Recipe}. The only difference between the two sources is the sign of \chieff, given in the legend.} \label{FigFit}
\end{figure}

A recipe for producing synthetic \chieff posteriors is as follows.
\begin{enumerate}
\item Generate a value for the median from the desired astrophysical distribution. This can be assumed to be the same as the true value, or can be offset from it using the values in Appendix~\ref{Sec.StatTables}.
\item Solve Eq.~(\ref{eq:skewness}) numerically for $\kappa$. The relevant $\gamma$ can be read from Fig.~\ref{FigSkew}.
\item Solve Eq.~(\ref{eq:variance}) for $\alpha$. This equation depends on both $\sigma$ and $\kappa$.
The $\sigma$ of each true \chieff obtained from our MCMC runs are given in Appendix~\ref{Sec.StatTables}. 
\item Use $\xi$, $\alpha$ and $\kappa$ so determined with Eqs.~(\ref{eq:GND})~and~(\ref{eq:why}) to get $p(\chieff)$.
\end{enumerate}
An illustrative example is shown in Fig.~\ref{FigFit} for two posteriors (histograms) and the corresponding synthetic version obtained with the method described above (lines).

In order to make use of these posteriors with alternative prior assumptions, the next step is to divide out the prior probability distribution for \chieff{} used in our injection and recovery.
The prior on \chieff{} is nontrivial, and combines the priors on the components of the spins $\boldsymbol{\chi}_i \cdot \mathbf{\hat L}$, and on the component masses.
The former assumes that the spins are isotropic in direction and that their dimensionless magnitudes are uniform between zero and $0.89$, a restriction chosen as the limit where the accelerated, reduced-order-quadrature likelihood we use is validated over~\citep{2016PhRvD..94d4031S}.
The prior on the aligned spin components has the simple analytic form given an upper magnitude $\chi_*$,
\begin{align}
p(\boldsymbol{\chi}_i \cdot \mathbf{\hat L}) = \frac{\ln(|\chi_*/\chi|)}{2 \chi_*} \,.
\end{align}
Meanwhile, the mass priors are uniform in the $m_1-m_2$ plane but constrained to lie within a range of constant $\mathcal M$ and with additional bounds on the maximum mass and minimum mass ratio~\citep{2016PhRvD..94d4031S}.

We find that we can fit the normalized prior distribution for \chieff{} with an analytic ansatz dependent on a single width parameter $w$.
The model is based on the double-exponential distribution, further constrained to require that $\chieff{}$ fall smoothly to zero at the maximum possible value $\chi_*$, which is a property inherited from the prior on $\mathbf{\chi}_i \cdot \mathbf{\hat L}$.
Our functional form is
\begin{align}
p(\chieff) = \frac{1 - e^{-(|\chieff| - \chi_*)/w}}{2\chi_* + 2 w (1 - e^{\chi_*/w})} \,.
\end{align}
Fitting to each prior individually yields a mean value $w = 0.23$ with a root-mean-square deviation $\sigma_w = 0.005$. 
We find that $w = 0.23$ gives an adequate fit to all the \chieff{} priors used in this study.

A webpage where the readers can generate synthetic \chieff posterior distributions (or likelihood) for the masses, spins and SNRs used in this paper can be found at \engineurl.

\section{Offsets from reduced-order quadratures}\label{Sec.ROQ}

As mentioned in the main body, we have used a ROQ approximation to the likelihood~\citep{2016PhRvD..94d4031S} implemented within the stochastic parameter estimation code \linf.
By expressing the overlap between the gravitational-wave data and the model waveform using a reduced basis, a ROQ likelihood can achieve speed-up factors between $\sim 10$--$300$ of a generic parameter estimation analysis.
The basis applicable for low-mass analyses (e.g. neutron star binaries) has recently been found to suffer from a issue~\citep{ROQbias} which, for some combination of masses, spins and SNRs, results in biased posteriors for the chirp mass and other intrinsic parameters, including~\chieff. No appreciable bias is observed in the extrinsic parameters (e.g. distance).
Thus, the original posteriors we obtained for \chieff for the BNS runs were biased away from their true value.

To correct for this bias we reanalyzed the BNS systems for which $\chieff=0$ (all three mass ratios) using the same waveform family as before, but without enabling the ROQ likelihood. 
Not using the ROQ likelihood causes the analysis to become significantly more computationally expensive, which is why we only ran the sources with $\chieff=0$, taking them to be representative for all BNS spins.
The reanalysis gave posteriors that, while not perfectly centered at the true value, were significantly closer to it. 
We have thus calculated the shift between the \chieff median of the ROQ and the non-ROQ runs, and applied those shifts to all the BNS runs, obtaining the points shown in Fig.~\ref{FigOffset}.

We note that BBH systems analyzed with the ROQ method are \emph{not} observed to be affected by this issue. This implies that none of the results published by the LIGO and Virgo collaborations suffer from this issue, since the ROQ method was not used to measure the mass and spins of GW170817~\citep{2017PhRvL.119p1101A}.

\section{Useful tables}\label{Sec.StatTables}

In this section we report key statistics for the posterior distributions of \chieff obtained with the simulations described in Sec.~\ref{Sec.Individual}.

\begin{table*}[!ht]
\caption{90\% credible intervals for BBHs at network SNR of 15. The BHs' tilt angles at 20 Hz are $10^\circ$.}\label{Tab.CIBBH_SNR15_tilt10}
\centering
\resizebox{\textwidth}{!}{
\begin{tabular}{l *{15}c}
\hline\hline
\multicolumn{16}{c}{\chieff}\\
\hline
$m_1-m_2$~[\msun] &-0.7 & -0.6 & -0.5 & -0.4 & -0.3 & -0.2 & -0.1 & 0 & 0.1 & 0.2 & 0.3 & 0.4 & 0.5 & 0.6 & 0.7 \\
\hline
30-30  & 0.351 & 0.344 & 0.334 & 0.319 & 0.307 & 0.286 & 0.247 & 0.227 & 0.226 & 0.216 & 0.203 & 0.183 & 0.172 & 0.159 & 0.149 \\
30-15  & 0.471 & 0.506 & 0.499 & 0.484 & 0.423 & 0.355 & 0.294 & 0.237 & 0.218 & 0.195 & 0.175 & 0.156 & 0.135 & 0.111 & 0.097 \\
30-10  & 0.627 & 0.617 & 0.588 & 0.549 & 0.510 & 0.449 & 0.408 & 0.378 & 0.322 & 0.292 & 0.275 & 0.241 & 0.191 & 0.135 & 0.088 \\
15-15  & 0.659 & 0.613 & 0.533 & 0.447 & 0.375 & 0.286 & 0.232 & 0.179 & 0.171 & 0.150 & 0.136 & 0.122 & 0.104 & 0.092 & 0.074 \\
15-7.5 & 0.603 & 0.633 & 0.568 & 0.587 & 0.525 & 0.442 & 0.354 & 0.278 & 0.241 & 0.215 & 0.185 & 0.165 & 0.146 & 0.107 & 0.073 \\
15-5   & 0.562 & 0.534 & 0.576 & 0.542 & 0.540 & 0.516 & 0.539 & 0.412 & 0.333 & 0.289 & 0.235 & 0.232 & 0.182 & 0.146 & 0.089 \\
5-5    & 0.548 & 0.522 & 0.499 & 0.427 & 0.362 & 0.292 & 0.238 & 0.191 & 0.181 & 0.161 & 0.164 & 0.138 & 0.122 & 0.100 & 0.062 \\
\hline\hline
\end{tabular}
}
\end{table*}

\begin{table*}[!ht]
\caption{90\% credible intervals for BBHs at network SNR of 30. The BHs' tilt angles at 20 Hz are $10^\circ$.}\label{Tab.CIBBH_SNR30_tilt10}
\centering
\resizebox{\textwidth}{!}{
\begin{tabular}{l *{15}c}
\hline\hline
\multicolumn{16}{c}{\chieff}\\
\hline
$m_1-m_2$~[\msun] &-0.7 & -0.6 & -0.5 & -0.4 & -0.3 & -0.2 & -0.1 & 0 & 0.1 & 0.2 & 0.3 & 0.4 & 0.5 & 0.6 & 0.7 \\
\hline\hline
30-30  & 0.205 & 0.189 & 0.183 & 0.171 & 0.165 & 0.153 & 0.143 & 0.121 & 0.122 & 0.112 & 0.104 & 0.093 & 0.088 & 0.083 & 0.078 \\
30-15  & 0.219 & 0.231 & 0.222 & 0.240 & 0.222 & 0.212 & 0.192 & 0.146 & 0.141 & 0.121 & 0.106 & 0.088 & 0.071 & 0.056 & 0.050 \\
30-10  & 0.257 & 0.276 & 0.305 & 0.326 & 0.325 & 0.282 & 0.210 & 0.182 & 0.187 & 0.185 & 0.160 & 0.128 & 0.088 & 0.064 & 0.045 \\
15-15  & 0.253 & 0.226 & 0.200 & 0.180 & 0.180 & 0.168 & 0.127 & 0.093 & 0.090 & 0.079 & 0.066 & 0.059 & 0.049 & 0.043 & 0.038 \\
15-7.5 & 0.367 & 0.466 & 0.506 & 0.518 & 0.441 & 0.353 & 0.272 & 0.210 & 0.169 & 0.140 & 0.131 & 0.120 & 0.093 & 0.064 & 0.043 \\
15-5   & 0.520 & 0.467 & 0.512 & 0.543 & 0.461 & 0.397 & 0.407 & 0.334 & 0.313 & 0.266 & 0.211 & 0.212 & 0.163 & 0.122 & 0.067 \\
5-5    & 0.263 & 0.265 & 0.277 & 0.238 & 0.231 & 0.208 & 0.167 & 0.141 & 0.131 & 0.113 & 0.103 & 0.089 & 0.081 & 0.074 & 0.048 \\
\hline\hline
\end{tabular}
}
\end{table*}

\begin{table*}[!ht]
\caption{90\% credible intervals for BBHs at network SNR of 15. The BHs' tilt angles at 20 Hz are $30^\circ$.}\label{Tab.CIBBH_SNR15_tilt30}
\centering
\resizebox{\textwidth}{!}{
\begin{tabular}{l *{15}c}
\hline\hline
\multicolumn{16}{c}{\chieff}\\
\hline
$m_1-m_2$~[\msun] &-0.7 & -0.6 & -0.5 & -0.4 & -0.3 & -0.2 & -0.1 & 0 & 0.1 & 0.2 & 0.3 & 0.4 & 0.5 & 0.6 & 0.7 \\
\hline
30-30  & 0.346 & 0.343 & 0.328 & 0.318 & 0.304 & 0.282 & 0.247 & 0.228 & 0.225 & 0.220 & 0.199 & 0.189 & 0.169 & 0.160 & 0.148 \\
30-15  & 0.407 & 0.441 & 0.477 & 0.467 & 0.428 & 0.365 & 0.292 & 0.239 & 0.216 & 0.194 & 0.171 & 0.153 & 0.135 & 0.111 & 0.096 \\
30-10  & 0.524 & 0.567 & 0.557 & 0.545 & 0.500 & 0.479 & 0.413 & 0.381 & 0.324 & 0.287 & 0.271 & 0.239 & 0.183 & 0.122 & 0.081 \\
15-15  & 0.676 & 0.598 & 0.533 & 0.447 & 0.360 & 0.283 & 0.210 & 0.185 & 0.163 & 0.156 & 0.133 & 0.118 & 0.104 & 0.089 & 0.075 \\
15-7.5 & 0.604 & 0.654 & 0.643 & 0.573 & 0.517 & 0.446 & 0.350 & 0.284 & 0.227 & 0.221 & 0.191 & 0.164 & 0.143 & 0.105 & 0.071 \\
15-5   & 0.491 & 0.576 & 0.566 & 0.528 & 0.468 & 0.516 & 0.525 & 0.427 & 0.344 & 0.289 & 0.249 & 0.235 & 0.178 & 0.140 & 0.081 \\
5-5    & 0.578 & 0.537 & 0.485 & 0.425 & 0.354 & 0.294 & 0.220 & 0.208 & 0.190 & 0.180 & 0.166 & 0.146 & 0.131 & 0.100 & 0.062 \\
\hline\hline
\end{tabular}
}
\end{table*}
 
\begin{table*}[!ht]
\caption{90\% credible intervals for BBHs at network SNR of 30. The BHs' tilt angles at 20 Hz are $30^\circ$.}\label{Tab.CIBBH_SNR30_tilt30}
\centering
\resizebox{\textwidth}{!}{
\begin{tabular}{l *{15}c}
\hline\hline
\multicolumn{16}{c}{\chieff}\\
\hline
$m_1-m_2$~[\msun] &-0.7 & -0.6 & -0.5 & -0.4 & -0.3 & -0.2 & -0.1 & 0 & 0.1 & 0.2 & 0.3 & 0.4 & 0.5 & 0.6 & 0.7 \\
\hline
 30-30  & 0.204 & 0.195 & 0.184 & 0.168 & 0.162 & 0.152 & 0.141 & 0.120 & 0.119 & 0.114 & 0.102 & 0.095 & 0.088 & 0.085 & 0.078 \\
 30-15  & 0.220 & 0.219 & 0.219 & 0.232 & 0.235 & 0.216 & 0.186 & 0.147 & 0.138 & 0.124 & 0.101 & 0.087 & 0.072 & 0.055 & 0.049 \\
 30-10  & 0.324 & 0.292 & 0.298 & 0.321 & 0.333 & 0.287 & 0.227 & 0.176 & 0.186 & 0.176 & 0.168 & 0.125 & 0.081 & 0.055 & 0.043 \\
 15-15  & 0.230 & 0.201 & 0.193 & 0.183 & 0.183 & 0.164 & 0.121 & 0.097 & 0.085 & 0.076 & 0.066 & 0.057 & 0.051 & 0.043 & 0.038 \\
 15-7.5 & 0.390 & 0.452 & 0.505 & 0.483 & 0.412 & 0.338 & 0.260 & 0.214 & 0.183 & 0.158 & 0.132 & 0.115 & 0.094 & 0.063 & 0.041 \\
 15-5   & 0.346 & 0.391 & 0.459 & 0.473 & 0.425 & 0.371 & 0.374 & 0.357 & 0.310 & 0.260 & 0.243 & 0.203 & 0.157 & 0.111 & 0.055 \\
 5-5    & 0.246 & 0.262 & 0.267 & 0.244 & 0.238 & 0.209 & 0.170 & 0.136 & 0.125 & 0.102 & 0.099 & 0.092 & 0.085 & 0.071 & 0.044 \\
\hline\hline
\end{tabular}
}
\end{table*}

\begin{table*}[!ht]
\caption{90\% credible intervals for BNSs at network SNR of 15. The spins are aligned with the orbital angular momentum.}\label{Tab.CIBNS_SNR15_tilt0}
\centering
\begin{tabular}{l *{9}c}
\hline\hline
\multicolumn{10}{c}{\chieff}\\
\hline
$m_1-m_2$~[\msun] &-0.2 & -0.15 & -0.1 & -0.05 & 0 & 0.05 & 0.1 & 0.15 & 0.2 \\
\hline
 1.4-1.4 & 0.237 & 0.204 & 0.184 & 0.152 & 0.138 & 0.136 & 0.126 & 0.120 & 0.110 \\
 2.0-1.4 & 0.272 & 0.241 & 0.214 & 0.185 & 0.167 & 0.153 & 0.139 & 0.142 & 0.129 \\
 2.2-1.3 & 0.309 & 0.270 & 0.239 & 0.204 & 0.184 & 0.167 & 0.152 & 0.139 & 0.135 \\
\hline\hline
\end{tabular}
\end{table*}

\begin{table*}[!ht]
\caption{90\% credible intervals for BNSs at network SNR of 30. The spins are aligned with the orbital angular momentum.}\label{Tab.CIBNS_SNR30_tilt0}
\centering
\begin{tabular}{l *{9}c}
\hline\hline
\multicolumn{10}{c}{\chieff}\\
\hline
$m_1-m_2$~[\msun] &-0.2 & -0.15 & -0.1 & -0.05 & 0 & 0.05 & 0.1 & 0.15 & 0.2 \\
\hline
 1.4-1.4 & 0.089 & 0.088 & 0.081 & 0.072 & 0.059 & 0.056 & 0.054 & 0.053 & 0.051 \\
 2.0-1.4 & 0.122 & 0.115 & 0.110 & 0.097 & 0.081 & 0.076 & 0.073 & 0.068 & 0.066 \\
 2.2-1.3 & 0.148 & 0.144 & 0.130 & 0.118 & 0.097 & 0.082 & 0.077 & 0.072 & 0.072 \\
\hline\hline
\end{tabular}
\end{table*}

\begin{table*}[!ht]
\caption{Standard deviation for BBHs at network SNR of 15. The BHs' tilt angles at 20 Hz are $10^\circ$.}\label{Tab.SigmaBBH_SNR15_tilt10}
\centering
\resizebox{\textwidth}{!}{
\begin{tabular}{l *{15}c}
\hline\hline
\multicolumn{16}{c}{\chieff}\\
\hline
$m_1-m_2$~[\msun] &-0.7 & -0.6 & -0.5 & -0.4 & -0.3 & -0.2 & -0.1 & 0 & 0.1 & 0.2 & 0.3 & 0.4 & 0.5 & 0.6 & 0.7 \\
\hline
 30-30  & 0.106 & 0.104 & 0.102 & 0.098 & 0.094 & 0.087 & 0.075 & 0.068 & 0.069 & 0.066 & 0.062 & 0.056 & 0.053 & 0.048 & 0.045 \\
 30-15  & 0.140 & 0.150 & 0.149 & 0.147 & 0.131 & 0.111 & 0.092 & 0.072 & 0.067 & 0.059 & 0.053 & 0.047 & 0.041 & 0.034 & 0.029 \\
 30-10  & 0.199 & 0.199 & 0.189 & 0.177 & 0.160 & 0.136 & 0.121 & 0.113 & 0.097 & 0.091 & 0.086 & 0.076 & 0.060 & 0.041 & 0.027 \\
 15-15  & 0.216 & 0.204 & 0.181 & 0.147 & 0.123 & 0.093 & 0.072 & 0.057 & 0.054 & 0.047 & 0.043 & 0.038 & 0.032 & 0.028 & 0.023 \\
 15-7.5 & 0.195 & 0.207 & 0.178 & 0.181 & 0.163 & 0.141 & 0.114 & 0.090 & 0.077 & 0.069 & 0.058 & 0.052 & 0.045 & 0.032 & 0.022 \\
 15-5   & 0.180 & 0.166 & 0.180 & 0.166 & 0.161 & 0.154 & 0.161 & 0.132 & 0.105 & 0.093 & 0.076 & 0.073 & 0.057 & 0.046 & 0.027 \\
 5-5    & 0.163 & 0.162 & 0.165 & 0.142 & 0.120 & 0.096 & 0.077 & 0.064 & 0.058 & 0.054 & 0.053 & 0.044 & 0.038 & 0.031 & 0.019 \\
\hline\hline
\end{tabular}
}
\end{table*}

\begin{table*}[!ht]
\caption{Standard deviation for BBHs at network SNR of 30. The BHs' tilt angles at 20 Hz are $10^\circ$.}\label{Tab.SigmaBBH_SNR30_tilt10}
\centering
\resizebox{\textwidth}{!}{
\begin{tabular}{l *{15}c}
\hline\hline
\multicolumn{16}{c}{\chieff}\\
\hline
$m_1-m_2$~[\msun] &-0.7 & -0.6 & -0.5 & -0.4 & -0.3 & -0.2 & -0.1 & 0 & 0.1 & 0.2 & 0.3 & 0.4 & 0.5 & 0.6 & 0.7 \\
\hline
 30-30  & 0.062 & 0.058 & 0.056 & 0.052 & 0.049 & 0.047 & 0.044 & 0.037 & 0.037 & 0.034 & 0.032 & 0.029 & 0.027 & 0.025 & 0.024 \\
 30-15  & 0.067 & 0.069 & 0.067 & 0.072 & 0.067 & 0.063 & 0.059 & 0.044 & 0.043 & 0.037 & 0.032 & 0.027 & 0.022 & 0.017 & 0.015 \\
 30-10  & 0.078 & 0.083 & 0.094 & 0.099 & 0.098 & 0.086 & 0.065 & 0.055 & 0.057 & 0.057 & 0.049 & 0.039 & 0.027 & 0.020 & 0.014 \\
 15-15  & 0.081 & 0.073 & 0.065 & 0.058 & 0.057 & 0.052 & 0.040 & 0.030 & 0.028 & 0.025 & 0.021 & 0.018 & 0.015 & 0.013 & 0.011 \\
 15-7.5 & 0.116 & 0.140 & 0.156 & 0.158 & 0.135 & 0.112 & 0.090 & 0.067 & 0.055 & 0.046 & 0.041 & 0.038 & 0.030 & 0.020 & 0.013 \\
 15-5   & 0.152 & 0.145 & 0.159 & 0.168 & 0.145 & 0.120 & 0.117 & 0.095 & 0.095 & 0.087 & 0.069 & 0.068 & 0.054 & 0.039 & 0.021 \\
 5-5    & 0.089 & 0.080 & 0.088 & 0.078 & 0.074 & 0.071 & 0.055 & 0.044 & 0.043 & 0.037 & 0.034 & 0.030 & 0.027 & 0.023 & 0.015 \\
\hline\hline
\end{tabular}
}
\end{table*}

\begin{table*}[!ht]
\caption{Standard deviation for BBHs at network SNR of 15. The BHs' tilt angles at 20 Hz are $30^\circ$.}\label{Tab.SigmaBBH_SNR15_tilt30}
\centering
\resizebox{\textwidth}{!}{
\begin{tabular}{l *{15}c}
\hline\hline
\multicolumn{16}{c}{\chieff}\\
\hline
$m_1-m_2$~[\msun] &-0.7 & -0.6 & -0.5 & -0.4 & -0.3 & -0.2 & -0.1 & 0 & 0.1 & 0.2 & 0.3 & 0.4 & 0.5 & 0.6 & 0.7 \\
\hline
 30-30  & 0.105 & 0.104 & 0.100 & 0.097 & 0.092 & 0.086 & 0.075 & 0.069 & 0.069 & 0.067 & 0.061 & 0.058 & 0.051 & 0.048 & 0.045 \\
 30-15  & 0.124 & 0.132 & 0.145 & 0.144 & 0.132 & 0.114 & 0.091 & 0.072 & 0.067 & 0.059 & 0.052 & 0.047 & 0.041 & 0.034 & 0.029 \\
 30-10  & 0.162 & 0.176 & 0.177 & 0.173 & 0.155 & 0.145 & 0.124 & 0.111 & 0.098 & 0.090 & 0.084 & 0.076 & 0.057 & 0.037 & 0.025 \\
 15-15  & 0.222 & 0.199 & 0.178 & 0.148 & 0.118 & 0.090 & 0.067 & 0.058 & 0.052 & 0.049 & 0.042 & 0.036 & 0.032 & 0.027 & 0.023 \\
 15-7.5 & 0.195 & 0.213 & 0.200 & 0.174 & 0.156 & 0.143 & 0.114 & 0.091 & 0.075 & 0.069 & 0.060 & 0.052 & 0.044 & 0.032 & 0.022 \\
 15-5   & 0.151 & 0.177 & 0.181 & 0.161 & 0.142 & 0.154 & 0.159 & 0.136 & 0.110 & 0.092 & 0.080 & 0.073 & 0.057 & 0.044 & 0.025 \\
 5-5    & 0.168 & 0.167 & 0.153 & 0.146 & 0.116 & 0.095 & 0.073 & 0.067 & 0.062 & 0.059 & 0.053 & 0.045 & 0.040 & 0.031 & 0.019 \\
\hline\hline
\end{tabular}
}
\end{table*}

\begin{table*}[!ht]
\caption{Standard deviation for BBHs at network SNR of 30. The BHs' tilt angles at 20 Hz are $30^\circ$.}\label{Tab.SigmaBBH_SNR30_tilt30}
\centering
\resizebox{\textwidth}{!}{
\begin{tabular}{l *{15}c}
\hline\hline
\multicolumn{16}{c}{\chieff}\\
\hline
$m_1-m_2$~[\msun] &-0.7 & -0.6 & -0.5 & -0.4 & -0.3 & -0.2 & -0.1 & 0 & 0.1 & 0.2 & 0.3 & 0.4 & 0.5 & 0.6 & 0.7 \\
\hline
 30-30  & 0.061 & 0.060 & 0.056 & 0.051 & 0.049 & 0.046 & 0.042 & 0.036 & 0.036 & 0.035 & 0.031 & 0.029 & 0.027 & 0.026 & 0.024 \\
 30-15  & 0.067 & 0.065 & 0.066 & 0.070 & 0.071 & 0.065 & 0.056 & 0.044 & 0.042 & 0.038 & 0.030 & 0.026 & 0.022 & 0.017 & 0.015 \\
 30-10  & 0.099 & 0.088 & 0.090 & 0.098 & 0.101 & 0.087 & 0.069 & 0.054 & 0.057 & 0.053 & 0.051 & 0.038 & 0.025 & 0.017 & 0.013 \\
 15-15  & 0.075 & 0.064 & 0.061 & 0.060 & 0.057 & 0.051 & 0.039 & 0.030 & 0.027 & 0.024 & 0.020 & 0.017 & 0.016 & 0.013 & 0.011 \\
 15-7.5 & 0.116 & 0.135 & 0.157 & 0.149 & 0.129 & 0.107 & 0.085 & 0.068 & 0.059 & 0.051 & 0.042 & 0.036 & 0.030 & 0.019 & 0.012 \\
 15-5   & 0.106 & 0.118 & 0.141 & 0.148 & 0.136 & 0.112 & 0.111 & 0.102 & 0.089 & 0.084 & 0.077 & 0.064 & 0.049 & 0.033 & 0.017 \\
 5-5    & 0.080 & 0.083 & 0.088 & 0.079 & 0.077 & 0.070 & 0.056 & 0.044 & 0.040 & 0.035 & 0.033 & 0.030 & 0.027 & 0.022 & 0.014 \\
\hline\hline
\end{tabular}
}
\end{table*}

\begin{table*}[!ht]
\caption{Standard deviation for BNSs at network SNR of 15. The spins are aligned with the orbital angular momentum.}\label{Tab.SigmaBNS_SNR15_tilt0}
\centering
\begin{tabular}{l *{9}c}
\hline\hline
\multicolumn{10}{c}{\chieff}\\
\hline
$m_1-m_2$~[\msun] &-0.2 & -0.15 & -0.1 & -0.05 & 0 & 0.05 & 0.1 & 0.15 & 0.2 \\
\hline
 1.4-1.4 & 0.078 & 0.067 & 0.060 & 0.051 & 0.046 & 0.044 & 0.040 & 0.039 & 0.036 \\
 2.0-1.4 & 0.091 & 0.081 & 0.072 & 0.061 & 0.054 & 0.050 & 0.047 & 0.047 & 0.042 \\
 2.2-1.3 & 0.103 & 0.090 & 0.079 & 0.068 & 0.060 & 0.054 & 0.051 & 0.046 & 0.045 \\
\hline\hline
\end{tabular}
\end{table*}

\begin{table*}[!ht]
\caption{Standard deviation for BNSs at network SNR of 30. The spins are aligned with the orbital angular momentum.}\label{Tab.SigmaBNS_SNR30_tilt0}
\centering
\begin{tabular}{l *{9}c}
\hline\hline
\multicolumn{10}{c}{\chieff}\\
\hline
$m_1-m_2$~[\msun] &-0.2 & -0.15 & -0.1 & -0.05 & 0 & 0.05 & 0.1 & 0.15 & 0.2 \\
\hline

 1.4-1.4 & 0.030 & 0.029 & 0.026 & 0.024 & 0.020 & 0.019 & 0.018 & 0.018 & 0.017 \\
 2.0-1.4 & 0.041 & 0.038 & 0.036 & 0.032 & 0.027 & 0.025 & 0.024 & 0.022 & 0.022 \\
 2.2-1.3 & 0.050 & 0.046 & 0.042 & 0.038 & 0.033 & 0.027 & 0.025 & 0.024 & 0.023 \\
\hline\hline
\end{tabular}
\end{table*}

\begin{table*}[!ht]
\caption{Skewness for BBHs at network SNR of 15. The BHs' tilt angles at 20 Hz are $10^\circ$.}\label{Tab.GammaBBH_SNR15_tilt10}
\centering
\resizebox{\textwidth}{!}{
\begin{tabular}{l *{15}c}
\hline\hline
\multicolumn{16}{c}{\chieff}\\
\hline
$m_1-m_2$~[\msun] &-0.7 & -0.6 & -0.5 & -0.4 & -0.3 & -0.2 & -0.1 & 0 & 0.1 & 0.2 & 0.3 & 0.4 & 0.5 & 0.6 & 0.7 \\
\hline
 30-30  & 0.100 & 0.119 & 0.065 & 0.085 & -0.052 & -0.120 & -0.213 & 0.004 & 0.077 & -0.073 & -0.150 & -0.120 & -0.110 & -0.065 & -0.085\\
 30-15  & 0.745 & 0.658 & 0.524 & 0.378 & 0.219 & 0.021 & 0.128 & 0.285 & 0.583 & 0.400 & 0.357 & 0.318 & 0.043 & -0.109 & -0.072       \\
 30-10  & 0.159 & 0.069 & -0.150 & -0.581 & -0.633 & -0.750 & -0.605 & -0.197 & 0.265 & 0.572 & 0.562 & 0.391 & 0.143 & -0.179 & -0.179 \\
 15-15  & 0.209 & 0.246 & 0.372 & 0.318 & 0.618 & 0.751 & 1.007 & 1.556 & 1.403 & 1.363 & 1.232 & 1.011 & 0.693 & 0.330 & -0.004        \\
 15-7.5 & -0.738 & -0.777 & -0.721 & -0.595 & -0.329 & 0.027 & 0.402 & 1.161 & 1.549 & 1.744 & 1.579 & 1.478 & 1.138 & 0.692 & 0.261    \\
 15-5   & -1.008 & -1.201 & -0.996 & -1.153 & -1.178 & -0.762 & -0.382 & 0.091 & 0.593 & 1.630 & 1.556 & 1.353 & 1.110 & 0.770 & 0.366  \\
 5-5    & 0.538 & 0.200 & 0.254 & 0.408 & 0.443 & 0.710 & 1.204 & 1.712 & 1.757 & 1.818 & 1.712 & 1.530 & 1.297 & 0.710 & 0.353         \\
\hline\hline
\end{tabular}
}
\end{table*}

\begin{table*}[!ht]
\caption{Skewness for BBHs at network SNR of 30. The BHs' tilt angles at 20 Hz are $10^\circ$.}\label{Tab.GammaBBH_SNR30_tilt10}
\centering
\resizebox{\textwidth}{!}{
\begin{tabular}{l *{15}c}
\hline\hline
\multicolumn{16}{c}{\chieff}\\
\hline
$m_1-m_2$~[\msun] &-0.7 & -0.6 & -0.5 & -0.4 & -0.3 & -0.2 & -0.1 & 0 & 0.1 & 0.2 & 0.3 & 0.4 & 0.5 & 0.6 & 0.7 \\
\hline
 30-30  & -0.021 & -0.002 & -0.146 & -0.087 & -0.082 & -0.032 & -0.143 & 0.001 & -0.120 & -0.050 & -0.076 & -0.101 & -0.047 & -0.008 & -0.057\\
 30-15  & -0.047 & -0.092 & 0.046 & 0.050 & 0.106 & -0.006 & -0.104 & -0.105 & 0.134 & 0.073 & 0.009 & 0.089 & -0.014 & -0.062 & 0.004       \\
 30-10  & 0.148 & 0.260 & 0.140 & 0.065 & -0.037 & -0.414 & -0.534 & -0.524 & 0.034 & -0.269 & -0.272 & -0.440 & -0.342 & -0.138 & 0.061     \\
 15-15  & 0.535 & 0.968 & 1.098 & 1.471 & 1.444 & 1.165 & 0.838 & 1.276 & 1.125 & 1.046 & 0.964 & 0.773 & 0.509 & 0.139 & 0.019              \\
 15-7.5 & 1.132 & 0.668 & 0.531 & 0.310 & -0.030 & -0.069 & 0.268 & 0.560 & 1.105 & 1.091 & 1.223 & 1.026 & 0.926 & 0.695 & 0.455            \\
 15-5   & 0.153 & 0.543 & 0.288 & -0.318 & -0.831 & -0.735 & -0.871 & -0.534 & 0.082 & 0.768 & 0.901 & 0.831 & 0.459 & 0.331 & 0.351         \\
 5-5    & 0.696 & 0.623 & 0.814 & 1.142 & 1.077 & 0.888 & 1.121 & 1.628 & 1.305 & 1.463 & 1.600 & 1.818 & 1.548 & 1.049 & 1.076              \\
\hline\hline
\end{tabular}
}
\end{table*}

\begin{table*}[!ht]
\caption{Skewness for BBHs at network SNR of 15. The BHs' tilt angles at 20 Hz are $30^\circ$.}\label{Tab.GammaBBH_SNR15_tilt30}
\centering
\resizebox{\textwidth}{!}{
\begin{tabular}{l *{15}c}
\hline\hline
\multicolumn{16}{c}{\chieff}\\
\hline
$m_1-m_2$~[\msun] &-0.7 & -0.6 & -0.5 & -0.4 & -0.3 & -0.2 & -0.1 & 0 & 0.1 & 0.2 & 0.3 & 0.4 & 0.5 & 0.6 & 0.7 \\
\hline
 30-30  & 0.052 & 0.071 & 0.092 & 0.037 & 0.036 & -0.139 & -0.233 & 0.001 & 0.062 & -0.122 & -0.138 & -0.129 & -0.098 & -0.077 & -0.133  \\
 30-15  & 0.609 & 0.611 & 0.517 & 0.363 & 0.157 & 0.155 & 0.094 & 0.338 & 0.548 & 0.463 & 0.350 & 0.235 & 0.047 & -0.104 & -0.046        \\
 30-10  & -0.133 & 0.057 & -0.166 & -0.393 & -0.672 & -0.774 & -0.570 & -0.249 & 0.251 & 0.629 & 0.628 & 0.481 & -0.002 & -0.135 & -0.194\\
 15-15  & 0.175 & 0.067 & 0.272 & 0.519 & 0.518 & 0.809 & 0.990 & 1.419 & 1.491 & 1.302 & 1.171 & 0.947 & 0.576 & 0.248 & 0.061          \\
 15-7.5 & -0.371 & -0.633 & -0.671 & -0.655 & -0.400 & 0.105 & 0.541 & 1.242 & 1.652 & 1.502 & 1.492 & 1.414 & 1.184 & 0.749 & 0.231     \\
 15-5   & -0.027 & -0.190 & -0.547 & -1.015 & -1.027 & -0.710 & -0.242 & 0.159 & 0.759 & 1.464 & 1.531 & 1.361 & 1.032 & 0.574 & 0.223   \\
 5-5    & 0.571 & 0.447 & 0.189 & 0.324 & 0.379 & 0.633 & 1.125 & 1.586 & 1.556 & 1.932 & 1.672 & 1.609 & 1.144 & 0.770 & 0.374          \\
\hline\hline
\end{tabular}
}
\end{table*}

\begin{table*}[!ht]
\caption{Skewness for BBHs at network SNR of 30. The BHs' tilt angles at 20 Hz are $30^\circ$.}\label{Tab.GammaBBH_SNR30_tilt30}
\centering
\resizebox{\textwidth}{!}{
\begin{tabular}{l *{15}c}
\hline\hline
\multicolumn{16}{c}{\chieff}\\
\hline
$m_1-m_2$~[\msun] &-0.7 & -0.6 & -0.5 & -0.4 & -0.3 & -0.2 & -0.1 & 0 & 0.1 & 0.2 & 0.3 & 0.4 & 0.5 & 0.6 & 0.7 \\
\hline
 30-30  & 0.056 & -0.115 & -0.091 & 0.003 & -0.008 & -0.135 & -0.077 & -0.128 & -0.060 & -0.068 & -0.072 & -0.067 & -0.049 & -0.039 & -0.076\\
 30-15  & -0.152 & -0.130 & 0.231 & 0.159 & 0.045 & -0.008 & -0.160 & -0.055 & 0.137 & 0.032 & 0.057 & -0.010 & 0.006 & 0.029 & 0.029       \\
 30-10  & 0.320 & 0.009 & 0.017 & 0.032 & -0.115 & -0.373 & -0.505 & -0.289 & -0.001 & -0.319 & -0.435 & -0.361 & -0.171 & -0.020 & 0.080   \\
 15-15  & 1.128 & 1.128 & 1.266 & 1.375 & 1.426 & 1.168 & 0.934 & 1.265 & 1.128 & 1.004 & 1.033 & 0.776 & 0.393 & 0.185 & -0.079            \\
 15-7.5 & 0.929 & 0.758 & 0.481 & 0.271 & 0.074 & -0.050 & 0.050 & 0.630 & 1.057 & 1.193 & 1.194 & 1.161 & 0.676 & 0.588 & 0.321            \\
 15-5   & 0.193 & 0.054 & 0.134 & -0.000 & -0.582 & -0.874 & -0.814 & -0.431 & 0.145 & 0.668 & 0.608 & 0.246 & 0.355 & 0.109 & 0.128        \\
 5-5    & 1.036 & 0.662 & 0.783 & 1.052 & 1.139 & 0.658 & 0.899 & 1.663 & 1.501 & 1.814 & 2.006 & 1.819 & 1.611 & 1.435 & 1.150             \\
\hline\hline
\end{tabular}
}
\end{table*}

\begin{table*}[!ht]
\caption{Skewness for BNSs at network SNR of 15. The spins are aligned with the orbital angular momentum.}\label{Tab.GammaBNS_SNR15_tilt0}
\centering
\begin{tabular}{l *{9}c}
\hline\hline
\multicolumn{10}{c}{\chieff}\\
\hline
$m_1-m_2$~[\msun] &-0.2 & -0.15 & -0.1 & -0.05 & 0 & 0.05 & 0.1 & 0.15 & 0.2 \\
\hline

 1.4-1.4 & 0.962 & 1.031 & 1.059 & 1.413 & 1.668 & 1.605 & 1.769 & 1.777 & 1.748\\
 2.0-1.4 & 0.787 & 0.911 & 1.107 & 1.250 & 1.540 & 1.705 & 1.684 & 1.889 & 1.831\\
 2.2-1.3 & 0.593 & 0.720 & 0.895 & 1.201 & 1.279 & 1.621 & 1.711 & 1.739 & 1.759\\
\hline\hline
\end{tabular}
\end{table*}

\begin{table*}[!ht]
\caption{Skewness for BNSs at network SNR of 30. The spins are aligned with the orbital angular momentum.}\label{Tab.GammaBNS_SNR30_tilt0}
\centering
\begin{tabular}{l *{9}c}
\hline\hline
\multicolumn{10}{c}{\chieff}\\
\hline
$m_1-m_2$~[\msun] &-0.2 & -0.15 & -0.1 & -0.05 & 0 & 0.05 & 0.1 & 0.15 & 0.2 \\
\hline

 1.4-1.4 & 1.996 & 1.837 & 1.798 & 1.814 & 1.961 & 2.181 & 1.938 & 2.065 & 2.040\\
 2.0-1.4 & 1.840 & 1.626 & 1.538 & 1.518 & 1.667 & 2.069 & 2.035 & 1.965 & 1.934\\
 2.2-1.3 & 1.803 & 1.474 & 1.417 & 1.330 & 1.424 & 1.830 & 1.902 & 1.951 & 1.708\\
\hline\hline
\end{tabular}
\end{table*}
\begin{table*}[!ht]
\caption{\chieffMedian for BBHs at network SNR of 15. The BHs' tilt angles at 20 Hz are $10^\circ$.}\label{Tab.BiasBBH_SNR15_tilt10}
\centering
\resizebox{\textwidth}{!}{
\begin{tabular}{l *{15}c}
\hline\hline
\multicolumn{16}{c}{\chieff}\\
\hline
$m_1-m_2$~[\msun] &-0.7 & -0.6 & -0.5 & -0.4 & -0.3 & -0.2 & -0.1 & 0 & 0.1 & 0.2 & 0.3 & 0.4 & 0.5 & 0.6 & 0.7 \\
\hline
 30-30  & -0.486 & -0.430 & -0.367 & -0.292 & -0.216 & -0.131 & -0.049 & 0.017 & 0.097 & 0.195 & 0.290 & 0.382 & 0.474 & 0.565 & 0.657   \\
 30-15  & -0.510 & -0.462 & -0.391 & -0.324 & -0.235 & -0.150 & -0.085  & -0.015 & 0.059 & 0.161 & 0.266 & 0.367 & 0.475 & 0.578 & 0.680 \\
 30-10  & -0.384 & -0.336 & -0.268 & -0.164 & -0.131 & -0.082 & -0.054  & -0.024 & 0.010 & 0.076 & 0.190 & 0.311 & 0.439 & 0.562 & 0.674 \\
 15-15  & -0.352 & -0.329 & -0.305 & -0.214 & -0.195 & -0.123 & -0.046 & 0.024 & 0.117 & 0.215 & 0.309 & 0.406 & 0.503 & 0.598 & 0.693   \\
 15-7.5 & -0.112 & -0.112 & -0.086 & -0.093 & -0.094 & -0.096 & -0.067  & -0.035 & 0.038 & 0.140 & 0.246 & 0.352 & 0.465 & 0.578 & 0.685 \\
 15-5   & -0.167 & -0.089 & -0.083 & -0.055 & -0.046 & -0.062 & -0.049  & -0.057 & -0.027 & 0.020 & 0.138 & 0.262 & 0.391 & 0.525 & 0.658\\
 5-5    & -0.443 & -0.321 & -0.280 & -0.237 & -0.170 & -0.103 & -0.050 & 0.028 & 0.120 & 0.216 & 0.313 & 0.411 & 0.511 & 0.616 & 0.711   \\
\hline\hline
\end{tabular}
}
\end{table*}

\begin{table*}[!ht]
\caption{\chieffMedian for BBHs at network SNR of 30. The BHs' tilt angles at 20 Hz are $10^\circ$.}\label{Tab.BiasBBH_SNR30_tilt10}
\centering
\resizebox{\textwidth}{!}{
\begin{tabular}{l *{15}c}
\hline\hline
\multicolumn{16}{c}{\chieff}\\
\hline
$m_1-m_2$~[\msun] &-0.7 & -0.6 & -0.5 & -0.4 & -0.3 & -0.2 & -0.1 & 0 & 0.1 & 0.2 & 0.3 & 0.4 & 0.5 & 0.6 & 0.7 \\
\hline
 30-30  & -0.629 & -0.549 & -0.459 & -0.373 & -0.279 & -0.182 & -0.083 & 0.004 & 0.098 & 0.195 & 0.294 & 0.392 & 0.489 & 0.585 & 0.683  \\
 30-15  & -0.618 & -0.556 & -0.476 & -0.392 & -0.293 & -0.192 & -0.095  & -0.006 & 0.073 & 0.180 & 0.281 & 0.380 & 0.484 & 0.584 & 0.688\\
 30-10  & -0.608 & -0.550 & -0.471 & -0.376 & -0.260 & -0.149 & -0.064  & -0.001 & 0.066 & 0.168 & 0.272 & 0.382 & 0.484 & 0.581 & 0.682\\
 15-15  & -0.595 & -0.540 & -0.449 & -0.368 & -0.273 & -0.172 & -0.073 & 0.013 & 0.109 & 0.206 & 0.302 & 0.399 & 0.497 & 0.596 & 0.694  \\
 15-7.5 & -0.492 & -0.488 & -0.383 & -0.323 & -0.199 & -0.140 & -0.095  & -0.023 & 0.040 & 0.144 & 0.249 & 0.354 & 0.464 & 0.575 & 0.684\\
 15-5   & -0.438 & -0.458 & -0.373 & -0.225 & -0.113 & -0.099 & -0.053  & -0.014 & 0.013 & 0.062 & 0.161 & 0.289 & 0.421 & 0.549 & 0.667\\
 5-5    & -0.613 & -0.480 & -0.417 & -0.351 & -0.242 & -0.156 & -0.066 & 0.017 & 0.119 & 0.213 & 0.311 & 0.404 & 0.506 & 0.607 & 0.704  \\
\hline\hline
\end{tabular}
}
\end{table*}

\begin{table*}[!ht]
\caption{\chieffMedian for BBHs at network SNR of 15. The BHs' tilt angles at 20 Hz are $30^\circ$.}\label{Tab.BiasBBH_SNR15_tilt30}
\centering
\resizebox{\textwidth}{!}{
\begin{tabular}{l *{15}c}
\hline\hline
\multicolumn{16}{c}{\chieff}\\
\hline
$m_1-m_2$~[\msun] &-0.7 & -0.6 & -0.5 & -0.4 & -0.3 & -0.2 & -0.1 & 0 & 0.1 & 0.2 & 0.3 & 0.4 & 0.5 & 0.6 & 0.7 \\
\hline
 30-30 & -0.470 & -0.419 & -0.366 & -0.293 & -0.219 & -0.133 & -0.050 & 0.019 & 0.095 & 0.193 & 0.290 & 0.385 & 0.478 & 0.567 & 0.655   \\
 30-15 & -0.487 & -0.451 & -0.389 & -0.307 & -0.230 & -0.155 & -0.080  & -0.017 & 0.054 & 0.158 & 0.266 & 0.370 & 0.475 & 0.580 & 0.686 \\
 30-10 & -0.276 & -0.316 & -0.262 & -0.199 & -0.126 & -0.081 & -0.055  & -0.024 & 0.010 & 0.074 & 0.185 & 0.304 & 0.453 & 0.575 & 0.685 \\
 15-15 & -0.358 & -0.281 & -0.270 & -0.252 & -0.186 & -0.125 & -0.047 & 0.026 & 0.118 & 0.215 & 0.310 & 0.407 & 0.505 & 0.600 & 0.693   \\
 15-7.5& -0.185 & -0.128 & -0.092 & -0.076 & -0.083 & -0.088 & -0.078  & -0.034 & 0.037 & 0.141 & 0.249 & 0.357 & 0.468 & 0.583 & 0.692 \\
 15-5  & -0.366 & -0.297 & -0.155 & -0.066 & -0.042 & -0.053 & -0.077  & -0.052 & -0.028 & 0.025 & 0.140 & 0.270 & 0.400 & 0.542 & 0.675\\
 5-5   & -0.463 & -0.377 & -0.264 & -0.236 & -0.155 & -0.095 & -0.040 & 0.028 & 0.125 & 0.219 & 0.315 & 0.410 & 0.516 & 0.616 & 0.711   \\
\hline\hline
\end{tabular}
}
\end{table*}

\begin{table*}[!ht]
\caption{\chieffMedian for BBHs at network SNR of 30. The BHs' tilt angles at 20 Hz are $30^\circ$.}\label{Tab.BiasBBH_SNR30_tilt30}
\centering
\resizebox{\textwidth}{!}{
\begin{tabular}{l *{15}c}
\hline\hline
\multicolumn{16}{c}{\chieff}\\
\hline
$m_1-m_2$~[\msun] &-0.7 & -0.6 & -0.5 & -0.4 & -0.3 & -0.2 & -0.1 & 0 & 0.1 & 0.2 & 0.3 & 0.4 & 0.5 & 0.6 & 0.7 \\
\hline
 30-30 & -0.611 & -0.537 & -0.454 & -0.371 & -0.282 & -0.184 & -0.086 & 0.004 & 0.098 & 0.196 & 0.296 & 0.395 & 0.491 & 0.589 & 0.684  \\
 30-15 & -0.573 & -0.526 & -0.467 & -0.389 & -0.294 & -0.191 & -0.090  & -0.006 & 0.079 & 0.183 & 0.282 & 0.384 & 0.484 & 0.587 & 0.695\\
 30-10 & -0.493 & -0.464 & -0.419 & -0.365 & -0.260 & -0.146 & -0.063  & -0.005 & 0.070 & 0.166 & 0.274 & 0.382 & 0.485 & 0.587 & 0.690\\
 15-15 & -0.644 & -0.546 & -0.454 & -0.368 & -0.271 & -0.172 & -0.075 & 0.014 & 0.109 & 0.206 & 0.302 & 0.401 & 0.500 & 0.597 & 0.697  \\
 15-7.5& -0.520 & -0.475 & -0.359 & -0.284 & -0.208 & -0.137 & -0.078  & -0.029 & 0.043 & 0.144 & 0.253 & 0.358 & 0.474 & 0.584 & 0.691\\
 15-5  & -0.498 & -0.451 & -0.383 & -0.267 & -0.118 & -0.091 & -0.058  & -0.020 & 0.018 & 0.058 & 0.197 & 0.332 & 0.437 & 0.570 & 0.684\\
 5-5   & -0.638 & -0.496 & -0.421 & -0.340 & -0.255 & -0.138 & -0.054 & 0.017 & 0.117 & 0.212 & 0.306 & 0.404 & 0.505 & 0.607 & 0.706  \\
\hline\hline
\end{tabular}
}
\end{table*}

\begin{table*}[!ht]
\caption{\chieffMedian for BNSs at network SNR of 15. The spins are aligned with the orbital angular momentum.}\label{Tab.BiasBNS_SNR15_tilt0}
\centering
\begin{tabular}{l *{9}c}
\hline\hline
\multicolumn{10}{c}{\chieff}\\
\hline
$m_1-m_2$~[\msun] &-0.2 & -0.15 & -0.1 & -0.05 & 0 & 0.05 & 0.1 & 0.15 & 0.2 \\
\hline

 1.4-1.4 & -0.144 & -0.101 & -0.057 & -0.016 & 0.024 & 0.071 & 0.118 & 0.167 & 0.216\\
 2.0-1.4 & -0.147 & -0.109 & -0.070 & -0.026 & 0.014 & 0.056 & 0.105 & 0.155 & 0.203\\
 2.2-1.3 & -0.144 & -0.113 & -0.076 & -0.039 & 0.006 & 0.045 & 0.093 & 0.142 & 0.194\\
\hline\hline
\end{tabular}
\end{table*}

\begin{table*}[!ht]
\caption{\chieffMedian for BNSs at network SNR of 30. The spins are aligned with the orbital angular momentum.}\label{Tab.BiasBNS_SNR30_tilt0}
\centering
\begin{tabular}{l *{9}c}
\hline\hline
\multicolumn{10}{c}{\chieff}\\
\hline
$m_1-m_2$~[\msun] &-0.2 & -0.15 & -0.1 & -0.05 & 0 & 0.05 & 0.1 & 0.15 & 0.2 \\
\hline

 1.4-1.4 & -0.187 & -0.135 & -0.086 & -0.037 & 0.012 & 0.060 & 0.109 & 0.158 & 0.207 \\
 2.0-1.4 & -0.205 & -0.153 & -0.102 & -0.051 & -0.003 & 0.044 & 0.094 & 0.145 & 0.195\\
 2.2-1.3 & -0.221 & -0.168 & -0.117 & -0.068 & -0.016 & 0.030 & 0.082 & 0.133 & 0.183\\
\hline\hline
\end{tabular}
\end{table*}

\clearpage \newpage %

\bibliography{spin_bias}

\end{document}